\documentclass[sn-mathphys,Numbered]{sn-jnl}% Math and Physical Sciences Reference Style
\usepackage{graphicx}%
\usepackage{multirow}%
\usepackage{soul}
\usepackage{amsmath,amssymb,amsfonts}%
\usepackage{amsthm}%
\usepackage{mathrsfs}%
\usepackage[title]{appendix}%
\usepackage{xcolor}%
\usepackage{textcomp}%
\usepackage{manyfoot}%
\usepackage{booktabs}%
\usepackage{algorithm}%
\usepackage{algorithmicx}%
\usepackage{algpseudocode}%
\usepackage{listings}%
\usepackage{xcolor}
\theoremstyle{thmstyleone}%
\theoremstyle{thmstyletwo}%

\theoremstyle{thmstylethree}%
\makeatletter
\renewcommand{\@maketitle}{\newpage\null%
    \if@remarkboxon\vbox to 0pt{\vspace*{-78pt}\hspace*{-18pt}\FMremark}\else\vskip21pt\fi%
    \hsize\textwidth\parindent0pt%
    {\hbox to \textwidth{\Artcatfont\ArtType\hfill}\par}
    \ifx\@title\empty\else%
        \removelastskip\vskip20pt\nointerlineskip%
        {\Titlefont\@title\par}
    \fi%
    \ifx\@subtitle\empty\else%
        \vskip9pt%
        {\SubTitlefont\@subtitle\par}
    \fi%
    \ifnum\aucount>0
        \global\punctcount\aucount%
        \vskip20pt%
        \artauthors\par%
        {\vskip7pt\addressfont\auaddress\par%
         \removelastskip\vskip24pt%
        \ifnum\emailcnt>0\relax%
           \ifx\corrauthemail\@empty\else{\ifnum\aucount>1*\fi}%
           Corresponding author(s). E-mail(s): \corrauthemail\par\fi%
           \ifx\authemail\@empty\else Contributing authors:\ \authemail\fi%
        \fi%
        \ifequalcont{\par$^{\dagger}$\@equalconttext\par}\fi%
         \removelastskip\vskip24pt%
        \ifpresentaddress{\par\@presentaddresstext\par}\fi%
        }
     \fi%
       \vspace{-20mm} % Adjust the -10mm to the amount of space you want to reduce
     {\printabstract\par} % Adjust the -10mm to the amount of space you want to reduce
     {\printkeywords\par}%
     \ifx\@pacs\empty\else%
       \loop\ifnum\PacsCount>0%
          \csname\romannumeral\PacsTmpCnt StorePacsTxt\endcsname\par%
          \StepDownCounter{\PacsCount}%
          \StepUpCounter{\PacsTmpCnt}%
       \repeat%
    \fi%
    \removelastskip\vskip36pt\vskip0pt}%
\makeatother

\definecolor{green}{RGB}{0,176,80}
\definecolor{blue}{RGB}{49,85,125}
\definecolor{red}{RGB}{103,0,31}
\definecolor{hl}{RGB}{255,0,0}

\begin{document}

\title[Article Title]{Timing the Escape of a Caged Electron}

\author[1]{\fnm{Connor} \sur{Fields}}

\author[2]{\fnm{Aleksandra} \sur{Foerster}}

\author[2]{\fnm{Sadegh} \sur{Ghaderzadeh}}

\author[2]{\fnm {Ilya} \sur{Popov}}

\author[2]{\fnm{Bang} \sur{Huynh}}

\author[1]{\fnm{Filipe} \sur{Junqueira}}

\author[1]{\fnm{Tyler} \sur{James}}

\author[1]{\fnm{Sofia} \sur{Alonso Perez}}

\author[2,3]{\fnm{David A} \sur{Duncan}}

\author[3]{\fnm{Tien-Lin} \sur{Lee}}

\author[1]{\fnm{Yitao} \sur{Wang}}

\author[4]{\fnm{Sally} \sur{Bloodworth}}

\author[4]{\fnm{Gabriela} \sur{Hoffman}}

\author[4]{\fnm{Mark} \sur{Walkey}}

\author[4]{\fnm{Richard J} \sur{Whitby}}

\author[4]{\fnm{Malcolm H} \sur{Levitt}}

\author[1]{\fnm{Brian} \sur{Kiraly}}

\author[1]{\fnm{James N} \sur{O'Shea}}

\author[2]{\fnm{Elena} \sur{Besley}}

\author[1]{\fnm{Philip} \sur{Moriarty}}

\affil[1]{\orgdiv{School of Physics \& Astronomy}, \orgname{University of Nottingham}, \orgaddress{\street{University Park}, \city{Nottingham}, \postcode{NG7 2RD}, \country{UK}}}

\affil[2]{\orgdiv{School of Chemistry}, \orgname{University of Nottingham},\orgaddress{\street{University Park}, \city{Nottingham}, \postcode{NG7 2RD}, \country{UK}}}

\affil[3]{\orgdiv{Diamond Light Source}, \orgname{Harwell Science \& Innovation Campus}, \orgaddress{\city{Didcot}, \postcode{OX11 0QX}, \country{UK}}}

\affil[4]{\orgdiv{School of Chemistry and Chemical Engineering}, \orgname{University of Southampton},\orgaddress{ \city{Southampton}, \postcode{SO17 1BJ}, \country{UK}}} 
%TC:ignore
\abstract{Charge transfer is fundamentally dependent on the overlap of the orbitals comprising the transport pathway. This has key implications for molecular, nanoscale, and quantum technologies, for which delocalization (and decoherence) rates are essential figures of merit. Here, we apply the core hole clock technique -- an energy-domain variant of ultrafast spectroscopy -- to probe the delocalization of a photoexcited electron inside a closed molecular cage, namely the Ar 2$p^54s^1$ state of Ar@C$_{60}$. Despite marginal frontier orbital mixing in the ground configuration, almost 80\% of the excited state density is found outside the buckyball due to the formation of a markedly diffuse hybrid orbital. Far from isolating the intracage excitation, the surrounding fullerene is instead a remarkably efficient conduit for electron transfer: we measure characteristic delocalization times of 6.6 $\pm$ 0.3 fs and $\lesssim$ 500 attoseconds, respectively, for a 3D Ar@C$_{60}$ film and a 2D monolayer on Ag(111).}
\pacs{79.20.Fv,  73.50.-h, 71.20.Tx, 82.30.Fi, 31.15.E} 
\keywords{endofullerene, core hole clock, charge transfer, X-ray standing wave, density functional theory, Auger-Meitner electron spectroscopy}
%TC:endignore

\maketitle

In an influential and enduring paper\cite{Hoffman1971}, Roald Hoffmann laid out a set of core principles associated with the interaction of localised orbitals in molecular systems, with a particular focus on the balance of through-space and through-bond coupling. Over fifty years later, Hoffmann's insights not only continue to underpin a great deal of what is now essentially seen as chemical ``intuition'', but multidisciplinary fields of research such as molecular electronics, photovoltaic/solar cell development (and photochemistry/photophysics in general), surface science, and nanoscience all owe a great deal to his work.   \\

\noindent Alongside what might be best described as the static coupling of orbitals explored by Hoffmann, a central focus of each of those fields -- molecular electronics in particular -- is the measurement, control, and exploitation of the tunnelling of carriers between, and through, units, contacts, and spacers in molecular and nanoscale architectures\cite{Worner2017}. In other words, it is the dynamic properties of charge delocalization and motion\cite{Wodtke2008,Peng2019,Sansone2010}, via mechanisms such as resonant, non-resonant, or superexchange tunnelling, thermally-dependent diffusive transport, and/or variable range hopping that are of especial interest \cite{Evers2020, Adams2003, Joachim2004}. These in turn determine the electrical conductance of a molecular or nanoscale component/junction, as described, for example, by the Landauer-Buttiker formalism (and subsequent modifications thereof) \cite{Landauer1988, Chen2021}. \\

\noindent We focus here on a molecular system that is unique in the context of through-space versus through-bond transport: endohedral fullerenes. Although their ``host-guest'' nature is of course not without chemical parallel\cite{Kfoury2018,CageReview,Ham2023}, no other chemical system -- including clathrates, inclusion complexes, zeolites, metal-organic frameworks, and supramolecular assemblies -- involves total encapsulation and containment inside a ``seamless'' framework, where the guest species cannot leave without covalent bonds being broken, as is the case for endofullerenes. This has critical implications in terms of the isolation of the encapsulate from its surrounding physicochemical environment and, as we shall see, for the dynamics of charge transfer to/from the encaged species.  \\

%-- clathrates, inclusion complexes, zeolites, metal-organic frameworks, and supramolecular assemblies each embed one component within another -- those examples all involve an ensemble of molecules. And while the host species of single-molecule-level complexes such as calixarenes, cryptophanes, and coordination cages each isolate, to a greater or lesser extent, their guest cargo,

\noindent In this context, and despite its chemical ``oddness'', Ar@C$_{60}$ --  a single argon atom encapsulated within a C$_{60}$ cage, Fig. 1(a) \cite{Saunders1994,Bloodworth2020Ar} -- is a particularly intriguing limiting case. In the ground state, there is remarkably little hybridisation of the encapsulated Ar with the frontier C$_{60}$ orbitals (i.e. highest occupied molecular orbital (HOMO), lowest unoccupied molecular orbital (LUMO), HOMO-1, LUMO+1 etc...) Although Morscher \textit{et al.}\cite{Morscher2010} provide compelling evidence for a hybrid Ar 3$p$-6$T_{1u}$ state, this is located 8 eV below the HOMO binding energy, i.e. $\sim$~10 eV below the Fermi level, and therefore well outside the energy range for electron transfer that underpins conductance in molecular electronics architectures. Given the marginal ground state coupling of the Ar atom with the fullerene frontier orbitals, one might ask whether this lack of overlap extends to excited states inside the cage. We have therefore measured the delocalization rate of a photoexcited state of the encapsulated argon. \\

\noindent Using the Auger resonant Raman variant\cite{Gelmukhanov1999} of the core hole clock technique\cite{Bjorneholm1992,Bruhwiler2002}, we monitor, with sub-femtosecond temporal resolution, the delocalization of a photoexcited Ar 4\textit{s} electron (Ar 2\textit{p}$_{3/2} \rightarrow 4s$) for Ar@C$_{60}$ molecules adsorbed as a bulk film or as a monolayer on a Ag(111) surface. For the latter, we complement the resonant Auger analysis with normal incidence X-ray standing wave (NIXSW) \cite{Woodruff2005} measurements, enabling, in parallel, an accurate determination of the position of the Ar atom above the substrate -- the first time, to the best of our knowledge, that the NIXSW technique has been combined with core hole clock analysis to enable a direct correlation of charge transfer rate with adsorbate geometry. \\

\noindent We find that the na\"ive picture of decoupled Ar and fullerene orbitals outlined above entirely fails to explain the electron delocalization dynamics that occur in the endofullerene system. Density functional theory (DFT) calculations combined with the maximum overlap method (MOM)\cite{Gilbert2008, Besley2021} reveal that the photoexcited state is exceptionally diffuse, with $\sim$ 80 \% of its density delocalised outside the cage. The hydrogenic superatom orbital (SAMO) states of fullerenes, first proposed by Feng \textit{et al.} \cite{Feng2008, Johansson2016}, are a compelling candidate for the origin of the extensive delocalisation. \\

\noindent Conceptually similar to, but distinct from, Rydberg orbitals, superatomic orbitals\cite{Feng2008,Johansson2016,Zhao2009} are not bound to the carbon atoms of the fullerene cage (unlike the traditional HOMO, LUMO etc...). Instead, SAMO states arise from the central potential of the core of the buckyball and are unique to hollow molecules; just as for the hydrogen atom, SAMO wavefunctions correspond to different orbital angular momentum states ($s$, $p$, $d$...) Of particular relevance to the interpretation of our core hole clock results, SAMO wavefunctions extend far beyond the carbon-atom-derived $\sigma$ and $\pi$ orbitals, to the extent that hybridization into metal-like nearly-free-electron bands occurs, with a substantial bandwidth ($\sim$ 600 meV) in the bulk fullerite crystal\cite{Zhao2009}. In the context of electron transfer, this represents a new and fascinating addition to Hoffmann's schema: interaction and delocalization via a coupling of atomic and superatomic orbitals. 

\section*{Results and Discussion}\label{sec2}

The core-hole clock (CHC) technique\cite{Bruhwiler2002, Menzel2008,Borges2019,Zharnikov2020}, first introduced in the early nineties\cite{Bjorneholm1992,Wurth1992}, is an energy-domain alternative to ultrafast pump-probe spectroscopy that is capable of measuring the rate of electron transfer on time scales ranging from tens of attoseconds\cite{Johansson2020} to $\sim$ 100 femtoseconds (depending on the lifetime, $\tau_{_{\mathrm{CH}}}$, of the particular core hole that is used as the clock\cite{Fohlisch2005}). CHC spectroscopy also has the key advantage of being chemically specific, with all of the attendant spectral ``fingerprinting'' advantages; this aspect is pivotal for the work described herein. \\ 

\noindent A  schematic of the CHC protocol used to determine the rate of delocalization of the photoexcited Ar 4\textit{s} state in Ar@C$_{60}$ is shown in Fig. 1(b). Resonant X-ray excitation from the Ar 2\textit{p}$_{3/2}$ level produces an initial core-excited Ar $2p^54s^1$ configuration.  There are then two primary channels for the subsequent decay of that excited state: (i) a spectator Auger-Meitner process, where the 4\textit{s} electron does not delocalize before decay of the core excitation, and (ii) ``traditional'' Auger-Meitner electron emission, where the photoexcited electron has tunnelled away from the original excitation site before core-hole decay. (Note that, as demonstrated by Fig. 2(c), the participator (or resonant photoemission) decay channel plays a negligible role in the case of Ar@C$_{60}$.) A key assumption here is that the core hole decay and electron delocalization rates are independent of each other. Moreover, both are assumed to follow a first order rate equation\cite{Bruhwiler2002,Menzel2008} and therefore decay with an exponential dependence. \\

\begin{figure}[t!]
\centering
\includegraphics[scale=0.42]{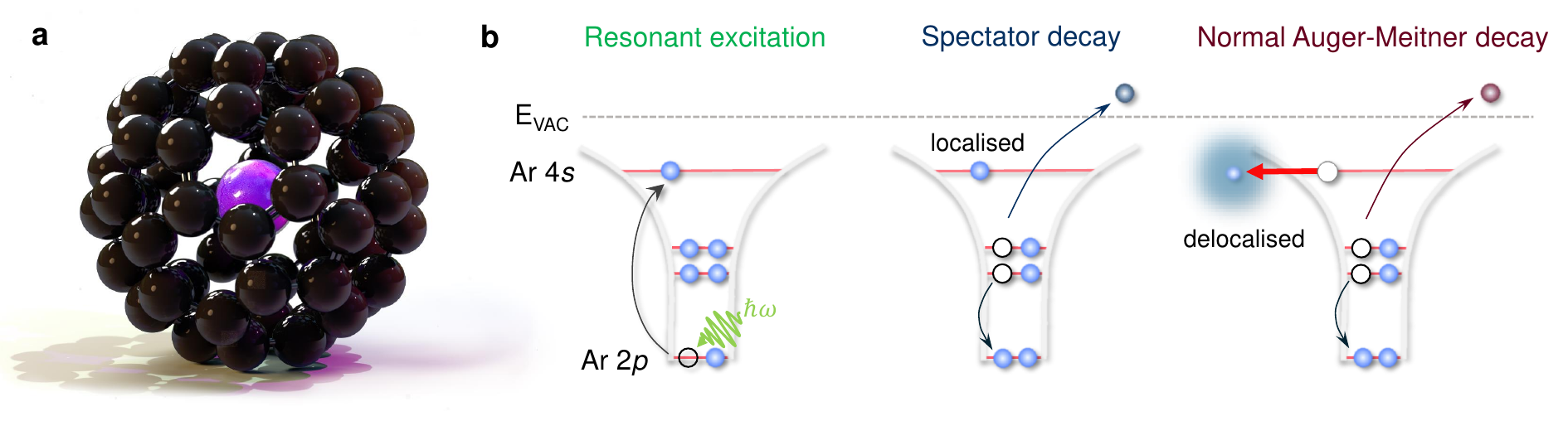}
\vspace{3mm}
\caption{\textbf{Core-hole clock spectroscopy of an endofullerene.
 (a)} 3D-rendered image of the ground state geometry of Ar@C$_{60}$ predicted by density functional theory (see \textit{Methods.}) \textbf{(b) The core-hole clock technique.} Following resonant excitation via X-ray absorption, the photo-excited Ar 2$p^5 4s^1$ state can decay either via a \textcolor{blue} {spectator Auger-Meitner process}, where the 4$s$ electron remains localized on the time scale of the core hole decay, or a \textcolor{red}{normal Auger-Meitner process}, for which the 4$s$ electron has tunnelled away (into the surrounding molecular matrix and/or substrate) before the core hole decays. The relative intensity of electron emission via these channels enables the delocalization rate of the 4$s$ state to be determined.}\label{<Fig1>}
\end{figure}

\begin{figure}[b!]
\centering
\includegraphics[scale=0.54]{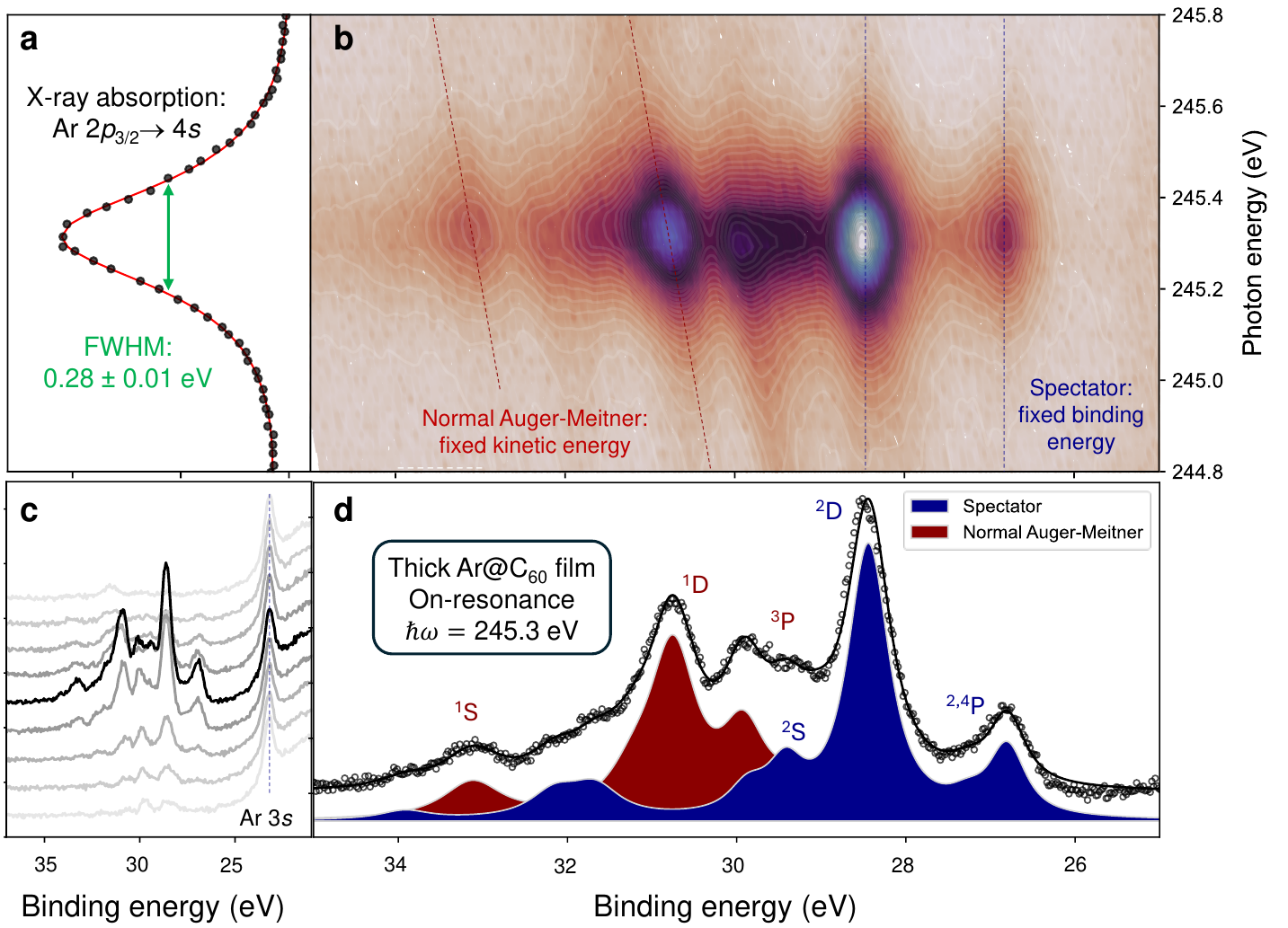}
\vspace{3mm}
\caption{\textbf{Timing electron delocalization in Ar@C$_{60}$.
 (a)} Ar $2p_{3/2} \rightarrow 4s$ partial electron yield X-ray absorption spectrum (filled circles) with Lorentzian fit (red line). \textbf{(b)} Resonant Auger-Meitner map showing intensity of decay spectra as a function of photon energy. Note that the X-ray absorption spectrum shown in (a) both shares its photon energy axis with the resonant Auger-Meitner map and is also the line-by-line integral of the map. (A total electron yield spectrum is included in the Supplementary Information (Fig. S2)). Normal Auger-Meitner transitions are at fixed kinetic energy and therefore disperse diagonally with photon energy (red dashed lines), whereas the spectator peaks are at fixed binding energy (blue dashed lines). For clarity, the dispersion (or lack thereof) of all peaks is not shown. \textbf{(c)} Auger-Meitner electron spectra spanning $\pm$400 mV either side of the resonance condition (black line), in 100 mV increments. The Ar 3\textit{s} peak intensity does not resonate and is constant to within 5\%, highlighting the negligible contribution of participator transitions to Ar core hole decay. (Note that the map in (b) and the set of spectra in (c) were acquired from different, but similarly prepared, Ar@C$_{60}$ samples.) \textbf{(d)}~On-resonance decay spectrum showing decomposition into the various normal Auger-Meitner and spectator components. Following Karis \textit{et al.}\cite{Karis1996}, we associate the shake-up features at $\sim$ 32 eV and $\sim$ 34 eV binding energy with spectator intensity of 3\textit{p}$^4$5\textit{s}$^{1}$ character. See S.I. for more detail on the fitting process. The map in (b) shares its binding energy axis with the on-resonance spectrum shown in (d).}\label{<Fig2>}
\end{figure}

\noindent We focus on X-ray absorption across the Ar 2$p_{3/2} \rightarrow 4s$ spectral peak. This is not only the initial excitation step in the CHC process (Fig. 1(b)), but the X-ray absorption spectrum by itself (Fig. 2(a)) already provides a great deal of insight into the degree of electronic coupling of the encapsulated Ar with the surrounding environment. The Ar 2$p_{3/2}\rightarrow 4s$ absorption spectrum shown in Fig. 2(a) (for a bulk film of Ar@C$_{60}$) is best fitted with a pure Lorentzian function, whose linewidth (full width at half-maximum (FWHM)) of 280 ($\pm 10$)~meV should be compared with the $\sim$~120~meV linewidth of gas phase argon\cite{Lizzit2009, Carroll2001}. Fig. 2(a) is a line-by-line integral of the resonant Auger-Meitner map of Fig. 2(b), i.e. it is a partial electron yield X-ray absorption measurement. (A fit to a total electron yield XAS spectrum (SI Fig. S2) results in a linewidth that agrees within experimental uncertainty, 260 $\pm$ 10 meV, with that of the partial yield spectrum.) \\

\noindent In the solid state, the extensive X-ray absorption, CHC, and photoemission measurements of argon on variously adsorbed graphene (Gr) monolayers reported by Lizzit \textit{et al.}\cite{Lizzit2013} arguably represent the most appropriate dataset with which to compare our Ar@C$_{60}$ XAS and CHC results. As described below, the Ar@C$_{60}$ system surprisingly exhibits behaviour at odds with that for the weakly coupled Gr/O/Ru, Gr/SiO$_2$, and Gr/SiC systems (i.e. unlike that expected for an isolated argon atom). \cite{Lizzit2013}.   \\

\subsection*{Quantifying the electron delocalisation time}
\noindent By decomposing the decay spectrum into its normal Auger-Meitner and spectator components (Fig. 2), the characteristic delocalization time (often simply called the charge transfer time), $\tau_{_{\mathrm{D}}}$, for the X-ray excited Ar 4s electron can be determined\cite{Menzel2008} (see \textit{Methods}, Figs. S3, S4, and the accompanying discussion in the supplementary information file). When adsorbed directly on a metal, the value of $\tau_{_{\mathrm{D}}}$ measured in this way for argon is of order a few fs\cite{Wurth2000, Martensson1995}; with a graphene monolayer sandwiched between the metal and argon, the on-resonance value of $\tau_D$ varies from $\sim$ 3 fs to 16 fs (depending on the level of graphene-metal interaction)\cite{Lizzit2013}; and for argon decoupled from the substrate via an underlying Ar/Xe spacer layer, the value of $\tau_{\mathrm{CT}}$ increases to over 50 fs\cite{Wurth2000}. (Indeed, the calculations of Gauyacq and Borisov\cite{Gauyacq2004} predict values of $\tau_D$ as large as 7 picoseconds for thick argon films.) In the context of the Ar@C$_{60}$ system where there is marginal mixing of the argon and fullerene density in the ground state, one might initially, and perhaps na\"ively, expect the charge transfer rate to be relatively slow -- comparable, at least, to that for the decoupled and weakly interacting Ar-on-graphene and Ar-on-Xe systems. (Although see the section titled \textit{Role of the Z+1 approximation} below.) This is not at all what we find.\\

\noindent  Despite the apparent chemical isolation of the encapsulated argon atom within the fullerene cage, the on-resonance value of $\tau_{_{\mathrm{D}}}$ for bulk Ar@C$_{60}$, 6.6$\pm$0.3 fs, shows that not only does electron delocalization occur on a time scale that is up to three orders of magnitude faster than that predicted for ``bare'' argon atoms condensed in a thick multilayer film (see, for example, Table II of Gauyacq and Borisov\cite{Gauyacq2004}) but that the charge transfer rate is comparable to that for argon separated from a metal substrate (namely Pt(111)) by a graphene monolayer\cite{Lizzit2013}, despite the Ar@C$_{60}$ solid having a band gap larger than 2~eV. Moreover, the primary trend of a reduction in delocalization time as a function of increasing photon energy (Fig. S4) is entirely opposite to that observed for argon adsorbed directly on a variety of metal surfaces (including Ag(111)), where the band structure of the substrate (and the concomitant wave-vector matching requirement) leads to larger values of $\tau_D$ as $\hbar\omega$ is increased\cite{Fohlisch2003}. \\

\noindent These observations all point to a substantial coupling and mixing of the core-excited argon 4$s$ state with the surrounding carbon cage, rather than an isolation of the excited state within the endofullerene. To interpret this mixing of the argon and fullerene density, and to gain a deeper understanding of the concomitant rapid transfer of the photoexcited 4$s$ electron, we turn to quantum chemistry calculations.

\subsection*{Beyond the confines of the cage: Ar 4$s$ delocalization} 
\noindent Despite the seeming lack of any interaction beyond dispersion forces in the endofullerene crystal (a van der Waals solid), there is clearly a relatively facile delocalization pathway available to the photoexcited Ar 4$s$ electron. Excited-state calculations exploiting the maximum overlap method (MOM)\cite{Gilbert2008,Besley2021} (see \textit{Methods} and S.I.) provide key insights into the rapid escape of the encaged Ar 4$s$ electron. (A justification of our use of the MOM, and a comparison with time-dependent density functional theory (TD-DFT) calculations, is given in the SI. We also discuss relativistic considerations at length in the SI.) \\

\noindent Fig. 3 shows isosurfaces and radial distribution functions for the ground-state and excited-state 4\textit{s} orbitals, with the latter calculated using the MOM. To estimate the spatial extent of the ground and photoexcited states, we have integrated the spherically averaged radial density distribution and determined the fraction of the density that is found at distances larger than the cage radius. For both the ground and excited states, the 4\textit{s} orbital extends significantly beyond the confines of the fullerene cage, with more than 80\% of the density lying beyond the Ar@C$_{60}$ radius of 3.54 \AA. However, while the ground state unoccupied 4\textit{s} orbital is of almost exclusively argon character (92\% contribution), the excited state instead has only a 13\% Ar contribution. In other words, the highly delocalised excited state is of majority carbon, i.e. fullerene cage, character.  (We use the C-squared population analysis method of Ros and Schuit\cite{Ros1966} to determine the contributions. See S.I. for a detailed discussion.) \\

\begin{figure}[b!]
\centering
\includegraphics[scale=0.55]{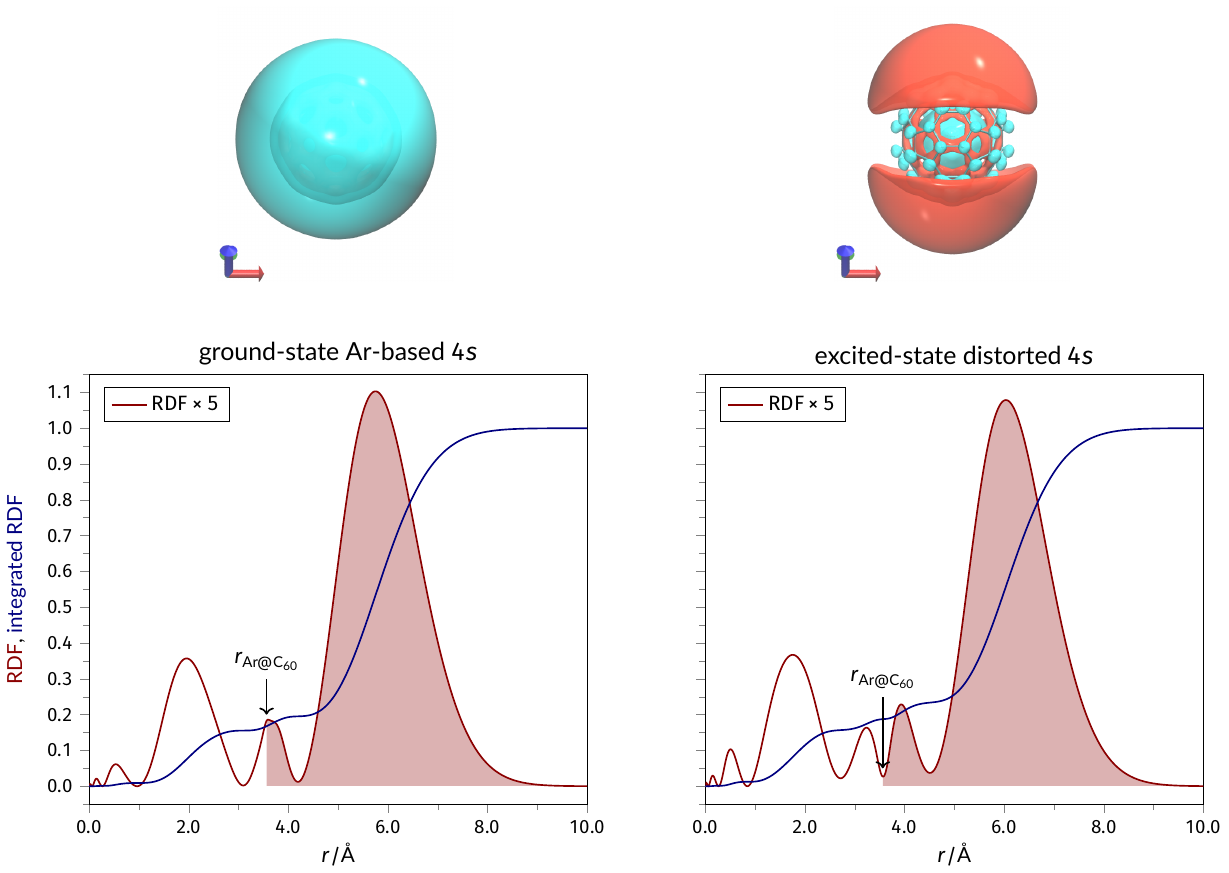}
\caption{\textbf{Ground vs core-excited 4$s$ state.} Isosurfaces and radial distribution functions for the ground-state and excited-state 4$s$ molecular orbitals. The isosurfaces are plotted at isovalues of $\pm$ 0.04 \AA$^{-3}$. The shaded area of each radial distribution function highlights the density that exists beyond the radius of the cage, $r_{Ar@C_{60}}$.} 
\label{<Fig3>}
\end{figure}

\noindent The combination of the dominant fullerene character and the highly diffuse nature of the excited state is characteristic of superatomic molecular orbitals\cite{Feng2008,Johansson2016,Zhao2009}. With this in mind, we have used the QSYM$^2$\cite{Huynh2024} framework to examine the symmetry of the excited-state orbital. Applying the relevant QSYM$^2$ projection operators (see SI for details), we find that the excited-state orbital shown in Fig. 3(b) comprises approximately 76\% \textit{S}-symmetry component, 23\% \textit{D}-symmetry component, and a very small contribution (less than 1\%) from $G$-symmetry. This is to be contrasted with the ground state 4$s$ orbital, which has essentially pure (i.e. almost 100\%) $S$-symmetry. The significant incorporation of the $D$-symmetry component in the excited 4$s$ orbital is attributed to the interaction with the carbon cage (as expected from the population analysis discussed above), providing a mechanism for mixing of the argon and fullerene density in a highly delocalised orbital.  \\   

\noindent Consideration of the relative energies of the levels underpinning the core hole clock experiment broadly supports the proposal of delocalization via mixing with SAMO density. By comparison with the Ar 2$p_{3/2}$ core level and HOMO binding energies, both referenced to the Fermi level ($E_{\mathrm{F}}$), we find that the Ar 2$p_{3/2} \rightarrow 4s$ resonance is located 4.80 $\pm$ 0.15 eV above the HOMO level (S.I., Fig. S1). This agrees remarkably well -- although see S.I. for a discussion of the role of core/valence excitons -- with the band structure calculations of Zhao \textit{et al.}\cite{Zhao2009}, which place the centre of the $s$-SAMO band at 4.8 eV above the HOMO level (i.e. resonant with the (core-excited) Ar 4$s$ energy). We also note that our measured value of the Ar 2$p \rightarrow 4s$ X-ray absorption resonance for $\sim$ 1 monolayer (ML) coverage (see below) of Ar@C$_{60}$ on Ag(111), \textit{viz.} 3.2 $\pm$ 0.1 eV above $E_{\mathrm{F}}$, is identical to that reported by Dutton \textit{et al.}\cite{Dutton2011} for the position of the $s$-SAMO resonance of the 1 ML C$_{60}$/Ag(111) system. Moreover, our measured on-resonance value of $\tau_D=6.5 \pm 0.1$ fs for the bulk Ar@C$_{60}$ film is entirely in line with the 4 -- 20 fs range for the s-SAMO lifetime (for empty C$_{60}$) determined by Zhu \textit{et al.} \cite{Zhu2006}. 

\subsection*{Role of the $Z$+1 approximation}
Argon in the core-excited 2$p^54s^1$ state is chemically very similar to ground state potassium: this is the well-known $Z+1$ approximation\cite{Jolly1970,Bagus2020} used extensively to interpret core level spectra and core hole clock measurements. Given that, in turn, potassium readily dopes the lowest unoccupied molecular orbital (LUMO) of C$_{60}$\cite{Hebard1991}, the rapid delocalisation of the photoexcited Ar 4$s$ electron that we observe could possibly arise from ``transient doping'' of the LUMO (although we highlight that, as discussed in the preceding paragraph, the Ar 4$s$ resonance lies significantly above the LUMO energy). We have investigated this ``K''-doping possibility at length using ground state DFT calculations.\\

\noindent We computed the ground state of K@C$_{60}$ at the PBE/6-31++G** level. Our results are very similar to those previously reported by \"Ostling and Ros\'en\cite{Ostling1993}. In particular, we find electron transfer from the encapsulated K atom to the C$_{60}$ LUMO, resulting in the occupation of one of the previously vacant $t_{1u}$ molecular orbitals and the close-to-complete (98\%) deoccupation of the K 4$s$ level (Fig. S9). However, in addition to the energy of the LUMO level being more than 2 eV below that of the Ar 4$s$ resonance in our X-ray absorption and core hole clock  measurements, we note that the spatial extent of the K-doped LUMO is considerably smaller than that of the photoexcited state shown in Fig. 3.\\

\noindent Moreover, and as described in more detail in the S.I., we find evidence for what we consider back-donation of electron density from the C$_{60}^-$ cage to the K$^+$ ion. Figures S9(a) and S9(b) show two occupied molecular orbitals that have significant mixing between potassium and the cage, and that also incidentally have $A_g$ symmetry in the $\mathcal{I}_h$ point group. This effect is noticeably absent in the excited state of Ar@C$_{60}$, where apart from the distorted 4$s$ molecular orbital, all occupied molecular orbitals reside either entirely on the Ar atom or entirely on the C$_{60}$ cage.

\subsection*{Separated, but connected: Ar@C$_{60}$/Ag(111)}
\noindent Arguably the most compelling experimental evidence for mixing of the photoexcited Ar 4$s$ state with the surrounding fullerene cage comes from our measurements of a chemisorbed monolayer of Ar@C$_{60}$ on Ag(111), in concert with ground state periodic projector augmented wave DFT (PAW-DFT) calculations (see \textit{Methods}.) We first focus on the measurement of the Ar atom position with respect to the Ag(111) surface via the X-ray standing wave technique (Fig. 4). NIXSW is an exceptionally powerful probe of adsorbate geometry\cite{Woodruff2005}, and is especially well-suited for endofullerene systems. Two key parameters result from an NIXSW measurement: the coherent fraction, $f_c$, a measure of the level of order in the adsorbate positions, and the coherent position, $p_c$ -- the position of the adsorbate with respect to the substrate scattering plane.\\ 

\begin{figure}[b!]
\centering
\includegraphics[scale=0.42]{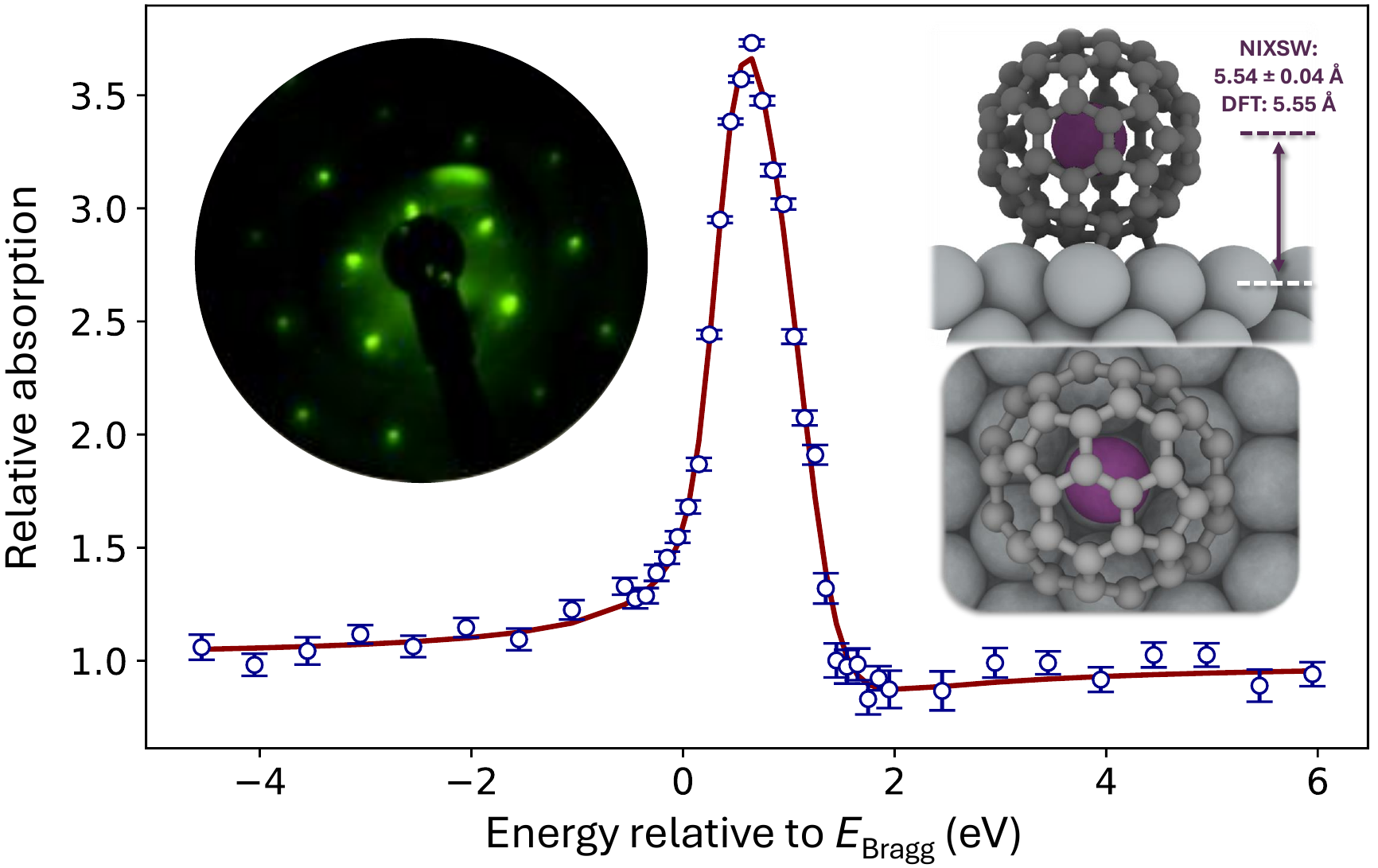}
\vspace{3mm}
\caption{\textbf{Locating the argon atom in adsorbed Ar@C$_{60}$.} The normal incidence X-ray standing wave profile derived from the variation in the Ar 2p$_{3/2}$ photoemission yield for an Ar@C$_{60}$ monolayer on Ag(111) is shown as the blue open circles in the main plot. A least squares fit to this profile (red line) (see \textit{Methods}) yields an Ar-Ag(111) separation of $5.54 \pm 0.04$ \AA, placing the Ar atom at the centre of the cage, despite the strong interaction of the surrounding fullerene with the Ag(111) surface. This separation is identical within experimental uncertainty to the Ar-Ag(111) adsorption height of 5.55 \AA$ $ predicted by our \textbf{(inset to right)} ground state PAW-DFT calculations for the 6:6 on-top geometry of the fullerene cage. \textbf{Inset to left:} ($2\sqrt{3} \times 2\sqrt{3}$)R$30^{o}$ LEED pattern for the Ar@C$_{60}$ monolayer.}
\end{figure}

\noindent Our deposition protocol (see \textit{Methods}) results in a value of $f_c$ for the encapsulated argon in the Ar@C$_{60}$ monolayer that is close to unity: 0.92$\pm$ 0.05, signifying a highly ordered molecular layer. The value of 5.54 $\pm$ 0.04 \AA\ for the argon atom height above the Ag(111) surface determined from the NIXSW analysis (Fig. 4) is identical to both the value predicted by our PAW-DFT calculations (5.55 \AA; see Fig. 4, \textit{Methods}, and S.I.) and the 5.5 $\pm 0.1$ \AA\ found by Pussi \textit{et al.}\cite{Pussi2012} (from a LEED analysis) for the centre of the (empty) C$_{60}$ cage in the metastable 6:6-bond-down, on-top adsorption geometry. Our monolayer preparation method very much favours adsorption in this kinetically limited state. As such, the experimentally measured 5.54 $\pm$ 0.04 \AA\ \mbox{Ar-Ag(111)} separation places the argon atom at the centre of the adsorbed endofullerene, its intracage position unperturbed by the chemisorption of the surrounding molecule.\\ 

 \begin{figure}[b!]
 \centering
 \includegraphics[scale=0.42]{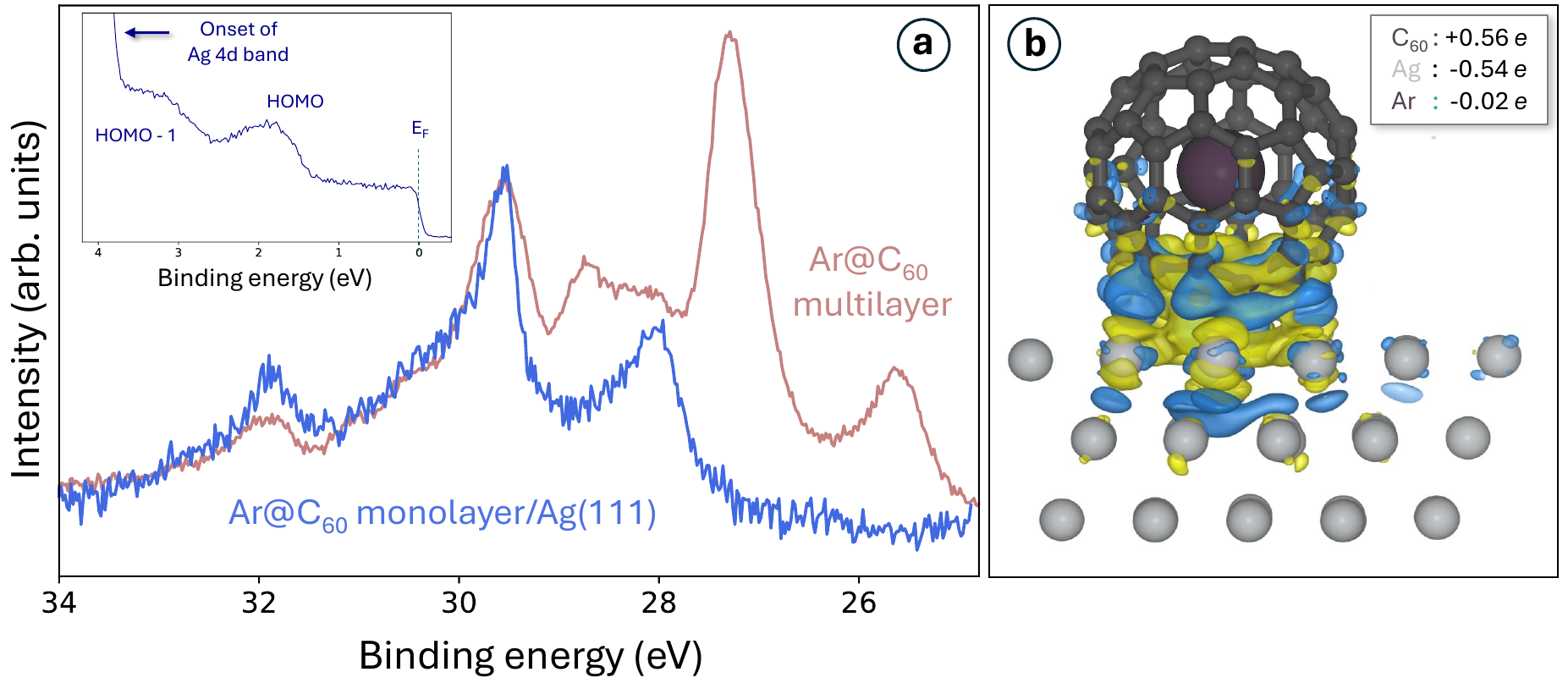}
 \vspace{3mm}
 \caption{\textbf{Escape in less than a femtosecond: charge transfer for a chemisorbed Ar@C$_{60}$ monolayer. }\textbf{(a)} On-resonance deexcitation spectrum (in blue) for Ar@C$_{60}$/Ag(111) following a linear background subtraction and a shift to 1.25 eV higher binding energy so as to align with the corresponding spectrum for the multilayer sample (purple). There is a complete absence of spectator peaks for the Ar@C$_{60}$/Ag(111) sample, and thus a sub-femtosecond Ar 4$s$ delocalization time. We estimate an upper limit of 500 attoseconds (see text). \textbf{Inset:} Valence band spectrum ($\hbar\omega=110$~eV) for the $\sim$ 1 ML Ar@C$_{60}$/Ag(111) sample. The HOMO and HOMO+1 binding energies exactly match those for 1 ML of empty C$_{60}$ on Ag(111)\cite{Pedio1999,Gibson2017}; \textbf{(b)} Ground state DFT calculation showing the difference in charge distribution for Ar@C$_{60}$ adsorbed on Ag(111), as compared to the isolated molecule and metal. The vast majority of the charge difference is restricted to the fullerene cage-Ag(111) interface; the ground state electron density of the Ar atom is almost entirely unaffected by adsorption. (Blue: depleted charge; yellow: gained charge. Isosurface: 7$\times 10^{-4} e$/\AA$^{-3}$.)}
 \end{figure}

\noindent For the metal-adsorbed Ar@C$_{60}$ monolayer, all trace of the spectator channel is removed and only the traditional Auger de-excitation pathway remains (Fig. 5(a)). The complete absence of spectator signal above the signal-to-noise ratio (SNR) limit of our experimental measurement means that the electron delocalization time is now at the sub-femtosecond level. Taking the magnitude of the measurement SNR into account\cite{Menzel2008}, we can place an upper limit on the value of $\tau_D$. We first determine the standard deviation, $\sigma$, of the background noise in the binding energy region (25 -- 27~eV) where we would expect spectator intensity to be located if it were present. Our criterion for signal detection is that the peak intensity should be a minimum of 3$\sigma$ above the background. On this basis, we find that the minimium detectable spectator signal would be a factor of 0.08 smaller than the normal Auger-Meitner intensity. (We note that this is very close to the factor of 0.1 estimated by F\"ohlisch \textit{et al.}\cite{Fohlisch2003} for the lowest practically resolvable charge transfer time available via the core hole clock method.) As such, we estimate that the upper limit of the electron delocalization time for the Ar@C$_{60}$ monolayer is $0.08\tau_{CH}$, i.e. $\sim$ 500 attoseconds. (Given the method of estimation, it is appropriate to quote only to 1 significant figure.) \\

\noindent This value is more than an order of magnitude smaller than for an argon atom adsorbed directly on the Ag(111) surface\cite{Fohlisch2003} (where the equilibrium adsorption height, namely 3.3~\AA \cite{Kirchner1994}, is more than 2 \aa ngstroms lower than that for argon in Ar@C$_{60}$). In other words, rather than acting to decouple the Ar 4$s$ excitation from the surrounding metallic environment (and thus impede the delocalisation rate), we instead see the same effect, now accentuated, as for the bulk Ar@C$_{60}$ film: the fullerene cage provides a remarkably efficient conduit for electron transfer. \\

\noindent Ground state PAW-DFT calculations (Fig. 5(b)) predict substantial charge transfer between the Ag(111) surface and the fullerene cage ($\sim 0.56 \textit{e}$, to be compared with the 0.5$e$ determined in a previous DFT study\cite{Wang2004} and 0.75$e$ estimated from photoemission measurements\cite{Tjeng1997}). However, the charge state of the encapsulated argon remains essentially unaffected by chemisorption of the fullerene cage. (The SI includes a discussion of the Ar 2$p$ and C 1$s$ photoemission spectra for the Ar@C$_{60}$ monolayer, which are consistent with this interpretation.)  It is clear, therefore, that the very significant enhancement of electron delocalization rate for the endofullerene monolayer again arises from excited state coupling of a diffuse Ar-fullerene hybrid orbital with the Ag(111) electronic structure, rather than an adsorption-induced shift in the position of the argon atom or its electronic structure. On this point, we note that the calculations of Gauyacq and Borisov\cite{Gauyacq2004} can be used to determine the charge delocalization time of an hypothetical ``bare'' argon atom adsorbed at the same height of 5.5~\AA \ as we measure for argon in Ar@C$_{60}$ on Ag(111). This is $\sim 65$~fs, at least two orders of magnitude slower  than the rate observed for the metal-adsorbed Ar@C$_{60}$ endofullerene. Moreover, SAMO-derived bands are robust against fullerene adsorption on metals\cite{Feng2008,Dutton2011} -- indeed, SAMO states were first observed in a C$_{60}$ monolayer on Cu(111)\cite{Feng2008} --  and so the absence of any spectator contribution to the de-excitation spectra of the Ar@C$_{60}$ monolayer is consistent with a coupling of an Ar-SAMO hybrid orbital to the electronic ``reservoir'' of the Ag(111) substrate. \\  

\section*{Conclusions.} The surprising result that emerges from our study is that encapsulating an inert atom in a closed carbon cage yields a substantially enhanced level of electronic coupling to the environment. We measure electron delocalization times that are at least an order of magnitude faster for Ar@C$_{60}$ than for a ``bare'' argon atom, despite the absence of ground state mixing of the frontier orbitals of the fullerene with the encaged argon. Our results are consistent with electron transport via diffuse hybrid Ar-fullerene orbitals, in which the vast majority of the electron density is found outside the cage. This has intriguing implications with regard to controlling the chemistry of endohedrally-caged atoms via delocalised hybrid orbitals. Adding submolecular spatial resolution to the XAS measurements via a strategy similar to that introduced by Ajayi \textit{et al.}\cite{Ajayi2023} is of particular future interest in this regard.            
%TC:ignore
\section*{Methods}\label{sec11}
\subsection*{Synthesis of Ar@C$_{60}$}
Ar@C$_{60}$ was synthesised by molecular surgery\cite{Bloodworth2022}, a process in which chemical reactions are used to open a hole in the C$_{60}$ cage large enough to allow argon to enter. A further series of reactions is then used to close the hole to reform the pristine C$_{60}$ cage, which now contains an argon atom\cite{Bloodworth2020Ar}. (Previously, Ar@C$_{60}$ has been obtained in very low yield by exposure of C$_{60}$ to argon at high temperatures and pressures followed by extensive purification (see, for example, Saunders \textit{et al.}\cite{Saunders1994}.)

\subsection*{Preparation of multilayer and monolayer films of Ar@C$_{60}$}
The Ag(111) surface was first cleaned via repeated sputter-anneal cycles (1 keV Ar$^+$ ions at an argon pressure of $\sim 2\times 10^{-5}$ mbar; sample annealing temperature $\sim 550^\mathrm{o}$C) until a sharp (1x1) low energy electron diffraction (LEED) pattern was visible and there was no evidence of C 1\textit{s} or O 1\textit{s} core-level signals in photoemission spectra (for which the photon energy was tuned to maximise the surface sensitivity of the photoelectrons.) Ar@C$_{60}$ was then deposited from a thermal evaporator operating at a temperature of 400 $(\pm 20) ^\mathrm{o}$C onto the Ag(111) sample, which was held at a temperature of $\sim$ 180 K throughout the deposition in order to prohibit reconstruction at the fullerene-Ag(111) interface\cite{Diehl2009}. This produced a ($2\sqrt{3} \times 2\sqrt{3}$)R$30^\mathrm{o}$ LEED pattern\cite{Diehl2009}.  Formation of monolayer coverages in this way essentially ``freezes out'' reconstruction (via ``nanopitting''\cite{Diehl2009,Pussi2012}) of the Ag(111) substrate, resulting in an X-ray standing wave coherent fraction value close to unity, substantially larger than that previously observed for endofullerene monolayers on Ag(111)\cite{Jarvis2021},  due to the homogeneity of the molecular adsorption sites. (Note, however, that a small ``overshoot'' in molecular coverage beyond the first monolayer is difficult to avoid using this protocol -- see S.I.) Multilayer coverages of sufficient thickness to quench photoemission signal from the Ag(111) substrate required cumulative deposition times of order four to five hours. A shift of $\sim$ 0.4 - 0.5 eV in the C 1$s$ core-level binding energy for monolayer vs multilayer coverages (see SI), coupled with a measurement of the ratio of the intensities of the C 1$s$ and Ag 3$d$ photoemission peaks, facilitated the identification of monolayer (and close-to-monolayer) coverage.  

\subsection*{Photoemission, X-ray absorption, and NIXSW measurements} All experimental work described in this paper was carried out at beamline I09, \textit{Surface and Interface Analysis}, at the Diamond Light Source\cite{LeeDuncan2018}. I09 is equipped with both a hard X-ray undulator, which was used for our NIXSW measurements, and a soft X-ray undulator, used for the acquisition of high-resolution C 1$s$, Ar 2$p$, Ag 3$d$, and valence band photoemission spectra, and for Ar $L_{2,3}$ and C K-edge X-ray absorption spectroscopy. (The resolving power of the soft X-ray branch is 10,000.) Considerable care was taken to reduce beam damage by detuning the beam (i.e. applying a small change in the undulator gap value to reduce peak intensity) and cooling the sample to temperatures between 100 K and 180 K. (See the Supplementary Information for more detail regarding the issue of beam damage.) The NIXSW data (Fig. 3) were acquired via the accumulation of Ar 2$p$ photoemission spectra during twelve separate sweeps of photon energy through the Ag(111) Bragg condition. (At 180 K the bulk lattice constant for Ag is 4.0779~\AA, equating to a (111) plane spacing of 2.354~\AA$ $ and a corresponding Bragg energy of 2633.47 eV). A Jupyter Notebook version of NIXSW/dynamical X-ray scattering code that had previously been developed by two of the authors (DAD and T-LL) was used to fit the data and extract the coherent position and coherent fraction parameters.   

\subsection*{Core hole clock considerations and fitting}
\noindent The determination of the delocalization/charge transfer time, $\tau_{_\mathrm{D}}$, is dependent on accurate knowledge of the core hole lifetime, $\tau_{CH}$. Our choice of Ar@C$_{60}$ (as opposed to other endofullerenes) for the measurement of intracage excited state delocalization was motivated in part by the ready availability of high precision measurements of the Ar 2p core-hole lifetime ($5.7 \pm 0.1$ fs\cite{Carroll2001}). Moreover, argon is a particularly attractive target species for CHC experiments due to the easily-resolved spectator shift, i.e. the difference in kinetic energy between the electron spectra arising from the two distinct decay channels in Fig. 1(b). The delocalization time, $\tau_D$, is calculated from the relative integrated intensities of the spectator and ``traditional'' Auger-Meitner contributions ($I_{\mathrm{Spec}}$ and $I_{\mathrm{Auger}}$, respectively) to the de-excitation spectra:  
\begin{equation}
\tau_{_{\mathrm{D}}}=\left(\frac{I_{\mathrm{Spec}}}{I_{\mathrm{Auger}}}\right)\tau_{_{\mathrm{CH}}}
\end{equation}
\noindent An additional motivation for the use of argon lies in the X-ray absorption linewidth, and, in particular, its relationship to the resolving power of the beamline ($\sim$ 10,000) at the Ar $L_3$ edge. As noted above, our measurements were acquired in the Auger-Meitner resonant Raman mode, for which the X-ray photon bandwidth ($\sim$ 25 meV in this case) is significantly smaller than the natural lifetime of the core-hole. When this is the case, the kinetic energy of the spectator peaks tracks the variation in photon energy across the absorption edge (for the reasons discussed by Menzel\cite{Menzel2008}.); in other words the spectator peaks remain at fixed binding energy (see Fig. 2). Conversely, the peaks arising from the ``traditional'' (i.e. non-Raman) Auger-Meitner process remain at fixed kinetic energy. These, and many other, constraints were applied during the fitting of the set of Auger-Meitner decay spectra acquired across the X-ray absorption resonance. An extensive description of our fitting strategy is given in the supplementary information.

\subsection*{Density functional theory (DFT) calculations.}
For the results shown in Fig.3, the structure of Ar@C$_{60}$ was first optimised at the DFT/PBE/6-31++G$^{**}$ level of theory and the orbitals involved in the Ar 2$p \rightarrow 4s$ transition were identified. The excited state was then calculated at the same level of theory, within the Q-Chem 5.4 package\cite{QChem}, with the aid of the maximum overlap method (MOM)\cite{Gilbert2008,Besley2021} (see following section) to maintain the core–hole during the DFT calculation. The ground state structural and charge transfer calculations of Fig. 3 were carried out with the Vienna Ab initio Simulation Package (VASP)~\cite{kresse1996efficient}, under periodic boundary conditions and within the plane-wave projector augmented-wave (PAW) method~\cite{kresse1999ultrasoft}. The Ag(111)-($2\sqrt{3} \times 2\sqrt{3}$)R30\textdegree-C$_{60}$ structures were optimized using the local spin density approximation (LSDA) with a force tolerance of $0.01$ eV\AA$^{-1}$ and an electronic convergence criterion of $10^{-6}$ eV. The energy cut-off was set to 500 eV, and a Monkhorst-Pack $k$-point grid of $3\times3\times1$ was used to sample the Brillouin zone. Electronic charges associated with individual atoms, utilized in the calculations of charge transfer, were derived using Bader analysis\cite{tang2009grid}. The atomic visualisations were generated using the Open Visualisation Tool (OVITO)\cite{Stukowski2009}. As described in the S.I., a variety of other DFT methods and adsorption geometries were employed to determine the Ar-Ag(111) separation but each provided poorer agreement with the NIXSW measurements than the LSDA approach. 

\subsection*{The maximum overlap method (MOM).} The MOM provides an efficient approach for calculating excited states by modifying the orbital selection step in the SCF procedure and targeting solutions with non-Aufbau occupations from a ground state reference set of molecular orbitals\cite{Gilbert2008,Barca2018}. By employing a simple orbital overlap-based criterion, the MOM prevents the variational collapse to the lowest energy solution. The MOM begins with an initial set of molecular orbitals (MOs) generated from the ground-state configuration of the system. Excitations are then introduced by modifying the occupation patterns, typically replacing one or more occupied orbitals with virtual orbitals. At each SCF iteration, the MOM algorithm applies an overlap metric to select the occupied orbitals that are most similar to the target orbitals from the previous iteration, guiding the SCF solver towards the intended excited state. In this work, we first calculated the electronic structure of Ar@C$_{60}$ at the PBE/6-31++G** level of theory to identify the relevant Ar-based $2p$ and $4s$ molecular orbitals.We then constructed an initial guess for the target excited state of this system by promoting an electron from one of the occupied Ar-based $2p$ molecular orbitals to the unoccupied Ar-based $4s$ molecular orbital. Subsequently, we used the MOM to relax the occupied molecular orbitals while staying as close as possible to the initial pattern.

%TC:endignore

\backmatter

\bmhead{Supplementary information} An extensive Supplementary Information file covering the following aspects of the work is available: circumventing beam damage; valence band photoemission for bulk Ar@C$_{60}$; comparison of Ar 2\textit{p} core-level photoemission and Ar 2\textit{p} $\rightarrow$ 4\textit{s} X-ray absorption spectra; fitting Auger-Meitner decay spectra; comparison of photoemission lineshapes for bulk and monolayer Ar@C$_{60}$ samples; Doniach-Sunjic asymmetry; raw on-resonance decay spectra for multilayer and monolayer Ar@C$_{60}$ coverage; population analysis: Ar and C contributions to the excited state; orbital symmetry decomposition; comparison with time-dependent DFT; relativistic considerations; Z+1 approximation and ground state K@C$_{60}$; and additional DFT calculations for determination of the Ar-Ag(111) separation.

\bmhead{Acknowledgments} We acknowledge very helpful discussions with Andrey Borisov and thank Dave McCue, Senior Beamline Technician, Diamond Light Source for his technical assistance and expertise.

\section*{Declarations}

\begin{itemize}
\item \textbf{Funding}. The synthesis of Ar@C$_{60}$  was supported by EPSRC (UK) (EP/M001962/1 and EP/P009980/1)), and the European Research Council (786707 – FunMagResBeacons). We gratefully acknowledge Diamond Light Source for the provision of beamtime via allocations SI-31574-1, SI31574-2, and SI-31574-3, and for the award of PhD studentship STU0425 (CF). PJM thanks the Engineering and Physical Sciences Research Council (EPSRC) for an Established Career Fellowship (EP/T033568/1). EB acknowledges a Royal Society Wolfson Fellowship. The computational work has been supported by the EPSRC Programme Grant ‘Metal Atoms on Surfaces and Interfaces (MASI) for Sustainable Future’ (EP/V000055/1)  and the University of Nottingham’s Augusta HPC service and the Sulis Tier 2 HPC platform funded by EPSRC Grant EP/ T022108/1 and the HPC Midlands+ consortium. DAD acknowledges a New Investigator Award from the Engineering and Physical Sciences Research Council [EP/X012883/1].  \\

\item \textbf{Conflicts of interest.} There are no conflicts of interest to declare.\\

\item \textbf{Consent for publication.} All authors have given consent for publication.\\

\item \textbf{Availability of data and materials.} All raw data is available at http://doi.org/10.17639/nott.7457\\

\item \textbf{Authors' contributions.} CF, FJ, TJ, SAP, DAD, YW, BK, JNOS, and PM carried out the experiments during three beamtime allocations at I09, Diamond Light Source; AF, SG, IP, BH, and EB were responsible for the DFT and MOM calculations; SB, GH, MW, MLH, and RJW provided the endofullerene samples; T-LL provided key advice and expertise related to the beamline experiments; CF and PM were responsible for experimental data analysis; PM drafted the paper, with input and feedback from all co-authors.     

\end{itemize}


\newpage

\begin{thebibliography}{99}
%TC:ignore
\bibitem{Hoffman1971}
Hoffman, R., Interaction of orbitals through space and through bonds, \textit{Acc. Chem. Res.} \textbf{4}, 1 (1971)
    
\bibitem{Worner2017}
W{\"o}rner, H. J., Arrell, C. A., Banerji, N., Cannizzo, A., Chergui, M., Das, A. K., Hamm, P., Keller, U., Kraus, P. M., Liberatore, E., \textit{et al.}, Charge migration and charge transfer in molecular systems, \textit{Struc. Dyn.} \textbf{4}, 061508 (2017)

\bibitem{Wodtke2008}
Wodtke, A. M., Matsiev, D., and Auerbach, D. J., 
Energy transfer and chemical dynamics at
solid surfaces: The special role of charge transfer,
\textit{Prog. Surf. Sci.} \textbf{83}, 167 (2008)

\bibitem{Peng2019} 
Peng P., Marceau C., and Villeneuve, D. M., Attosecond imaging of molecules using high harmonic spectroscopy, \textit{Nat. Rev. Phys.} \textbf{1}, 144 (2019)


\bibitem{Sansone2010}
Sansone, G. \textit{et al.}, Electron localization following attosecond molecular photoionization, \textit{Nature} \textbf{465}, 763 (2010)


\bibitem{Evers2020}
Evers, F., Koryt\'ar, R., Tewari, S.,  and van Ruitenbeek, J. M, Advances and challenges in single-molecule electron transport, \textit{Rev. Mod. Phys.} \textbf{92}, 035001 (2020)

%\bibitem{Nazmutdinov2023}
%Nazmutdinov, R. R., Shermokhamedov, S. A., Zinkicheva, T. T., Ulstrup, J., and Xiao X.,Understanding molecular and electrochemical charge transfer: theory and computations, \textit{Chem. Soc. Rev.}, \textbf{52}, 6230 (2023) 

\bibitem{Adams2003}
Adams, D. M. \textit{et al.}, Charge transfer on the nanoscale: current status, \textit{J. Phys. Chem. B} \textbf{107}, 6668 (2003)

%\bibitem{Naleway1991}
%Naleway, C. A., Curtiss, L. A., Miller, J. R., Superexchange-pathway model for long-distance electronic couplings, \textit{J. Phys. Chem.} \textbf{95}, 8434 (1991)

\bibitem{Joachim2004}
Joachim, C. and Ratner, M. A., Molecular wires: guiding the super-exchange interactions between two electrodes, \textit{Nanotechnology} \textbf{15}, 1065 (2004)

\bibitem{Landauer1988}
Landauer, R., Spatial variation of currents and fields due to localized scatterers in metallic conduction, \textit{IBM J. Res. Dev.} \textbf{32}, 306 (1988) and  \textit{IBM J. Res. Dev.} \textbf{1}, 223 (1957)

%\bibitem{Lambert2015}
%Lambert, C. Basic concepts of quantum interference and electron transport in single-molecule electronics. \textit{Chem. Soc. Rev.} \textbf{44} 875 (2015)

%\bibitem{Buttiker1986}
%Buttiker, M., Four-terminal phase-coherent conductance, \textit{Phys. Rev. Lett.} \textbf{57} 1761 (1986)

%\bibitem{Joachim2005}
%Joachim, C. and Ratner, M. A., Molecular electronics: Some views on transport junctions and beyond, \textit{Proc. Nat. Acad. Sci.} \textbf{102}, 8801 (2005)

\bibitem{Chen2021}
Chen, H. and Stoddart, J. F., From molecular to supramolecular electronics, \textit{Nature Rev. Mater.} \textbf{6} 804 (2021)

\bibitem{Kfoury2018}
Kfoury, M., Landy, D., and Fourmentin, S., Characterization of Cyclodextrin/Volatile Inclusion Complexes: A Review, \textit{Molecules} \textbf{23}, 1204 (2018)

\bibitem{Ham2023}
Ham, R., Nielsen, C. J., Pullen, S., and Reek, J. N. H.,Supramolecular Coordination Cages for Artificial Photosynthesis and Synthetic Photocatalysis, \textit{ Chem. Rev.} \textbf{123}, 5225 (2023)

\bibitem{CageReview} Mont\`a-Gonz\'alez, G., Sancen\'on, F., Mart\'inez-M\'a\~{n}ez, R., and Mart\'i-Centelles, V. Purely covalent molecular cages and containers for guest encapsulation, \textit{Chem. Rev.} \textbf{122}, 13636 (2022)

\bibitem{Saunders1994} Saunders, M., Jimenez-Vazquez, H. A, Cross, R. J., Mroczkowski, S., Gross, M. L., Giblin, D. E., Poreda, R. J., Incorporation of helium, neon, argon, krypton, and xenon into fullerenes using high pressure, \textit{J. Am. Chem. Soc.} \textbf{116}, 2193 (1994)

\bibitem{Bloodworth2020Ar} Bloodworth, S., Hoffman, G., Walkey, M. C.,  Bacanu, G. R., Herniman, J. M., Levitt, M. H., and Whitby, R. J., Synthesis of Ar@C$_{60}$ using molecular surgery, \textit{Chem. Commun}. \textbf{56}, 10521 (2020)



%\bibitem{Mandelcorn1959}
%Mandelcorn, L., Clathrates, \textit{Chem. Rev.} \textbf{59}, 827 (1959)

 

%\bibitem{Perez2022}
%P\'erez-Botella, E., Valencia, S., and Rey, F., Zeolites in Adsorption Processes: State of the Art and Future Prospects, \textit{Chem. Rev.} \textbf{122}, 17647 (2022)

%\bibitem{Griffin2022}
%Griffin, S. and Champness, N.R., A periodic table of metal-organic frameworks, \textit{Coord. Chem. Rev.} \textbf{414}, 213295 (2022)

%\bibitem{Goronzy2018}
%Goronzy, D. P., Ebrahimi, M., Rosei, F., Arramel, A., Yuan Fang, De Feyter, S., Tait, S.L., Chen Wang, Beton, P. H., Wee, A. T. S., Weiss, P. S., and Perepichka, D. F., Supramolecular Assemblies on Surfaces: Nanopatterning, Functionality, and Reactivity, \textit{ACS Nano} \textbf{12}, 7445 (2018)

%\bibitem{Bohmer1995}
%B\"ohmer, V., Calixarenes, Macrocycles with (Almost) Unlimited Possibilities, \textit{Angew. Chem. Int. Ed. Engl.} \textbf{34}, 713 (1995)

%\bibitem{Brotin2009} 
%Brotin, T. and Dutasta, J. -P., Cryptophanes and Their Complexes—Present and Future, \textit{Chem. Rev.} \textbf{109}, 88 (2009)


\bibitem{Morscher2010}
Morscher, M., Seitsonen, A. P., Ito, S., Takagi, H., Dragoe, N., and Greber, T., Strong $3p-T_{1u}$ hybridization in Ar@C$_{60}$, \textit{Phys. Rev. A.} \textbf{82}, 051201(R) (2010)

\bibitem{Gelmukhanov1999}
Gel'mukhanov, F. and Agren, H., Resonant X-ray Raman scattering, \textit{Phys. Rep.} \textbf{312}, 87 (1999)

\bibitem{Bjorneholm1992}
Bj\"orneholm, O., Nilsson, A., Sandell, A., Hernn\"as, B., and M\aa rtensson, N., Determination of time scales for charge transfer screening in physisorbed molecules, \textit{Phys. Rev. Lett.} \textbf{68}, 1892 (1992)

\bibitem{Bruhwiler2002}
Br\"uhwiler, P. A., Karis O., and Martensson, N., Charge-transfer dynamics studied using resonant core spectroscopies, \textit{Rev. Mod. Phys.} \textbf{74}, 703 (2002), and references therein.

\bibitem{Woodruff2005}
Woodruff,  D. P.,Surface structure determination using X-ray standing waves, \textit{Rep. Prog. Phys.} \textbf{68} 743 (2005); Zegenhagen, J., Surface structure determination with X-ray standing waves, \textit{Surf. Sci. Rep.} \textbf{18}, 202 (1993)

\bibitem{Gilbert2008}
Gilbert, A. T. B., Besley, N. A., and Gill, P. M. W., Self-consistent field calculations of excited states using the maximum overlap method (MOM), \textit{Phys. Chem. A} \textbf{112}, 13164 (2008)

%\bibitem{Besley2009}
%Besley, N. A., Gilbert, A. T. B., and Gill, P. M. W., Self-consistent-field calculations of core excited states, \textit{J. Chem. Phys.} \textbf{130}, 124308 (2009)

\bibitem{Besley2021}
Besley, N. A. Modeling of the spectroscopy of core electrons with density functional theory, \textit{WIREs Comput. Mol. Sci.}, \textbf{11} e1527 (2021)

\bibitem{Feng2008} 
Feng, M., Jin Zhao, and Petek, H., Atomlike, hollow-core–bound molecular orbitals of C$_{60}$, \textit{Science} \textbf{320}, 359 (2008)

\bibitem{Johansson2016}
Johansson, J. O., Bohl, E., and Campbell, E. E. B, Super-atom molecular orbital excited states of fullerenes, \textit{Phil. Trans. R. Soc. A} \textbf{374}, 20150322 (2016)

\bibitem{Zhao2009} Zhao, J., Feng M., and Petek, H. The superatom states of fullerenes and their hybridization into the nearly free electron bands of fullerenes. \textit{ACSNano} \textbf{3}, 853 (2009)

\bibitem{Menzel2008}
Menzel, D., Ultrafast charge transfer at surfaces accessed by core electron spectroscopies, \textit{Chem. Soc. Rev.} \textbf{37}, 2212 (2008)

\bibitem{Borges2019}
Borges, B. G. A. L., Roman, L. S., and Rocco, M. L. M., Femtosecond and attosecond electron transfer dynamics of semiconductors probed by core‑hole clock spectroscopy, \textit{Top. Cat.} \textbf{62}, 1004 (2019)

\bibitem{Zharnikov2020} Zharnikov, M. Femtosecond charge transfer dynamics in monomolecular films in the context of molecular electronics, \textit{Acc. Chem. Res.} \textbf{53}, 2975 (2020)

\bibitem{Wurth1992}
Wurth, W., Feulner, P., and Menzel, D., Resonant excitation and decay of adsorbate core holes, \textit{Physica Scripta} \textbf{T41}, 213 (1992)

\bibitem{Johansson2020}
Johansson, F. O. L. \textit{et al.}, Tailoring ultra-fast charge transfer in MoS$_2$, \textit{Phys. Chem. Chem. Phys.} \textbf{22}, 10335 (2020)

\bibitem{Fohlisch2005}
F\"ohlisch, A., Feulner, P., Hennies, F., Fink, A., Menzel, D., Sanchez-Portal, D., Echenique, P. M., and Wurth, W., Direct observation of electron dynamics in the attosecond domain, \textit{Nature} \textbf{436}, 373 (2005)

\bibitem{Lizzit2009}
Lizzit, S., Zampieri, G., Kostov, K. L., Tyuliev, G., Larciprete, R., Petaccia, L., Naydenov, B., and Menzel, D., Charge transfer from core-excited argon adsorbed on clean and hydrogenated Si(100): ultrashort timescales and energetic structure, \textit{New. J. Phys.} \textbf{11}, 053005 (2009)

\bibitem{Carroll2001}
Carroll, T. X., Bozek, J. D., Kukk, E., Myrseth, V., S\ae thre, L. J., and Thomas, T. D., Line shape and lifetime in argon 2p electron spectroscopy, \textit{J. Elec. Spec. Rel. Phen.} \textbf{120}, 67 (2001)

\bibitem{Lizzit2013} Lizzit, S., Larciprete, R., Lacovig, P., Kostov, K. L., and Menzel, D. Ultrafast charge transfer at monolayer graphene surfaces with varied substrate coupling, \textit{ACS Nano} \textbf{7}, 4359 (2013)

\bibitem{Wurth2000}
Wurth, W. and Menzel, D., Ultrafast electron dynamics at surfaces probed by resonant Auger spectroscopy, \textit{Chem. Phys.}\textbf{251}, 141 (2000)

\bibitem{Martensson1995}
M\aa rtensson, N. and Nilsson, A., Autoionization as a tool for studying adsorbed atoms and molecules, \textit{J. Elec. Spec. Rel. Phen.} \textbf{72},1 (1995)

\bibitem{Karis1996}
Karis, O., Nilsson, A., Weinelt, M., Wiell, T.,  Puglia, C., Wassdahl, N., M\aa rtensson, N., Samant, M., and St\"ohr, J., One-step and two-step description of deexcitation processes in weakly interacting systems, \textit{Phys. Rev. Lett.} \textbf{76}, 1380 (1996)

\bibitem{Gauyacq2004} Gauyacq, J. P. and Borisov, A. N., Excited electron transfer between a core-excited Ar$^*(2p_{3/2}^{-1}4s)$ atom and the metal substrate in the Ar/Cu(111) system, \textit{Phys. Rev. B} \textbf{69}, 235408 (2004)

\bibitem{Ros1966} Ros, P. and Schuit, G. C. A., Molecular orbital calculations on copper chloride complexes, \textit{Theor. Chim. Acta.} \textbf{4}, 1 (1966)

\bibitem{Huynh2024} Huynh, B. C., Wibowo-Teale, M, and Wibowo-Teale, A. M. QSYM$^2$: A quantum symbolic symmetry analysis program for electronic structure, \textit{J. Chem. Theor. Comp.} \textbf{20}, 114 (2024)

\bibitem{Jolly1970} Jolly, W. L. and Hendrickson, D. N. Thermodynamic interpretation of chemical shifts in core-electron binding energies, \textit{J. Am. Chem. Soc.} \textbf{92}, 1863 (1970)

\bibitem{Bagus2020} Bagus, P. S., Sousa, C.,  and Illas, F. Limitations of the equivalent core model for understanding core-level spectroscopies,  \textit{Phys. Chem. Chem. Phys.} \textbf{22}, 22617 (2020)

\bibitem{Hebard1991} Hebard, A. F., Rosseinsky, M. J., Haddon, R. C., Murphy, D. W., Glarum, S. H., Palstra, T. T. M., Ramirez, A. P, and Kortan, A. R. Superconductivity at 18 K in potassium-doped C$_{60}$, \textit{Nature} \textbf{350}, 600 (1991)

\bibitem{Ostling1993} \"Ostling, D. and Ros\'en, A. Electronic properties of the C$_{60}$ molecule doped with potassium, \textit{Chem. Phys. Lett.} \textbf{202}, 389 (1993)

\bibitem{Fohlisch2003}
F\"ohlisch, A \textit{et al.}, Energy dependence of resonant charge transfer from adsorbates to metal substrates, \textit{Chem. Phys.} \textbf{289}, 107 (2003)

\bibitem{Pedio1999}
Pedio M., Hevesi, K., Zema, N., Capozi, M., Perfetti, P., Gouttebaron, R., Pireaux, J. -J., Caudano, R., and Rudolf, P., C$_{60}$/metal surfaces: adsorption and decomposition, \textit{Surf. Sci.} \textbf{437}, 249 (1999)

\bibitem{Gibson2017}
Gibson, A. J., Temperton, R. H., Handrup, K., and O' Shea, J.N. Resonant core spectroscopies of the charge transfer interactions between C$_{60}$ and the surfaces of Au(111), Ag(111), Cu(111) and Pt(111). \textit{Surf. Sci.} \textbf{657}, 69 (2017)

\bibitem{Rubensson2000}
Rubensson, J. E., RIXS dynamics for beginners, \textit{J. Elec. Spec. Rel. Phen.} \textbf{110--111}, 135 (2000)

\bibitem{Arantes2013}
Arantes, C., Borges, B. G. A. L, Beck, B., Ara\'ujo, G., Roman, L. S., and Rocco, M. L. M., Femtosecond electron delocalization in poly(thiophene) probed by resonant Auger spectroscopy, J. Phys. Chem. C \textbf{117}, 8208 (2013)

%\bibitem{Johansson2020arxiv}
%Johansson,  F. O. L., Sassa, Y., Edvinsson, T., and Lindblad, A., Gone in 23 attoseconds: Charge transfer in resonantly core excited black phosphorous, \url{https://arxiv.org/pdf/2003.13394} (2020)



%\bibitem{Holmlin2001} Holmlin, R. E., Ismagilov, R. F., Haag, R.,
%Mujica, V.,Ratner, M. A., Rampi, M. A., and Whitesides, G. M., \textit{Angew. Chem. Int. Ed.} \textbf{40}, 2316 (2001)

%\bibitem{Koch2012} Koch, M., Ample, F., Joachim, C., and Grill, L., \textit{Nature Nanotech.} \textbf{7}, 713 (2012)

%\bibitem{Smalley1995} Smalley, J. F., Feldberg, S. W., Chidsey, C. E. D., Linford, M. R., Newton, M. D., and Liu, Y., \textit{J. Phys. Chem.} \textbf{99} 13141 (1995)

\bibitem{Dutton2011} Dutton, G. J., Dougherty, D. B., Jin, W., Reutt-Robey, J. E., and Robey, S. W., Superatom orbitals of C$_{60}$ on Ag(111): Two-photon photoemission and scanning tunneling spectroscopy,\textit{Phys. Rev. B} \textbf{84}, 195435 (2011) 

\bibitem{Zhu2006} Zhu, X. Y., Dutton, G., Quinn, D. P., Lindstrom, C. D., and Truhlar, D. G. Molecular quantum well at the C$_{60}$/Au(111) interface. \textit{Phys. Rev. B} \textbf{74}, 241401 (2006)



\bibitem{Pussi2012} Pussi, K., Li, H. I., Shin, H., Serkovic Loli, L. N., Shukla, A. K., Ledieu, J., Fournee, V., Wang, L. L., Su, S. Y., Marino, K. E., Snyder, M. V., and Diehl, R. D. Elucidating the dynamical equilibrium of C$_{60}$ molecules on Ag(111), \textit{Phys. Rev. B} \textbf{86}, 205406 (2012)

\bibitem{Kirchner1994} Kirchner, E. J. J., Kleyn, A. W., and Baerends, E.J. A comparative study of Ar/Ag(111) potentials. \textit{J. Chem. Phys.} \textbf{101}, 9155 (1994)

\bibitem{Wang2004} Wang, L. L., and Cheng, H.-P., Density functional study of the adsorption of C$_{60}$ on Ag(111) and Au(111) surfaces. \textit{Phys. Rev. B} \textbf{69}, 165417 (2004)

\bibitem{Tjeng1997} Tjeng, L. H., Hesper, R., Heessels, A. C. L., Heeres, A., Jonkman, H. T., and Sawatzky, G.A. Development of the electronic structure in a K-doped C$_{60}$ monolayer on a Ag(111) surface. \textit{Sol. St. Comm.} \textbf{103}, 31 (1997)



\bibitem{Ajayi2023} Ajayi, T.M., \textit{et al.}, Characterization of just one atom using synchrotron X-rays, \textit{Nature} \textbf{618}, 69 (2023)

\bibitem{Diehl2009}
Li, H. I, Pussi, K., Hanna, K. J., Wang, L-. L., Johnson, D. D., Cheng, H. -P., Shin, H., Curtarolo, S., Moritz, W., Smerdon, J. A., McGrath, R., and Diehl, R.D., Surface geometry of C$_{60}$ on Ag(111), \textit{Phys. Rev. Lett.} \textbf{103}, 056101 (2009)

\bibitem{Jarvis2021} Jarvis, S. P. \textit{et al.}, Chemical shielding of H$_2$O and HF encapsulated inside a C$_{60}$ cage, \textit{Comm. Chem.} \textbf{4}, 135 (2021)

\bibitem{LeeDuncan2018}
Lee, T.-L. and Duncan, D. A., A two-colour beamline for electron spectroscopies at Diamond Light Source, \textit{Synch. Rad. News} \textbf{31}, 16 (2018)

%\bibitem{Jolly1970}
%Jolly, W. L. and Hendrickson, D. N., Thermodynamic interpretation of chemical shifts in core-electron binding energies, \textit{J. Am. Chem. Soc.} \textbf{92}, 1863 (1970)

%\bibitem{Nyberg1999}
%Nyberg, M., Luo, Y., Triguero, L., Pettersson, L. G. M., and \AA gren, H., Core-hole effects in x-ray-absorption spectra of fullerenes, \textit{Phys. Rev. B} \textbf{60}, 7956 (1999)









%\bibitem{DS1970}
%Doniach S. and Sunjic M., Many-electron singularity in X-ray photoemission and X-ray line spectra from metals, \textit{J. Phys. C} \textbf{3}, 285 (1970)

%\bibitem{MeierNote} In a somewhat related vein, Meier \textit{et al.} (\textit{Nature Comm} \textbf{6} 8112 (2015)) have observed a similarly surprising dependence of the bulk dielectric constant of H$_2$O@C$_{60}$ on the nuclear spin (ortho vs para) of the encapsulated water molecule. 

\bibitem{Bloodworth2022}
Bloodworth, S. and Whitby R. J. Synthesis of endohedral fullerenes by molecular surgery. \textit{Commun. Chem.} \textbf{5}, 121 (2022).



% \bibitem{QChem} Y. Shao, Z. Gan, E. Epifanovsky, A. T. B. Gilbert, M. Wormit, J. Kussmann, A. W. Lange, A. Behn, J. Deng, X. Feng, D. Ghosh, M. Goldey P. R. Horn, L. D. Jacobson, I. Kaliman, R. Z. Khaliullin, T. Kús, A. Landau, J. Liu, E. I. Proynov, Y. M. Rhee, R. M. Richard, M. A. Rohrdanz, R. P. Steele, E. J. Sundstrom, H. L. Woodcock III, P. M. Zimmerman, D. Zuev, B. Albrecht, E. Alguire, B. Austin, G. J. O. Beran, Y. A. Bernard, E. Berquist, K. Brandhorst, K. B. Bravaya, S. T. Brown, D. Casanova, C.-M. Chang, Y. Chen, S. H. Chien, K. D. Closser, D. L. Crittenden, M. Diedenhofen, R. A. DiStasio Jr., H. Dop, A. D. Dutoi, R. G. Edgar, S. Fatehi, L. Fusti-Molnar, A. Ghysels, A. Golubeva-Zadorozhnaya, J. Gomes, M. W. D. Hanson-Heine, P. H. P. Harbach, A. W. Hauser, E. G. Hohenstein, Z. C. Holden, T.-C. Jagau, H. Ji, B. Kaduk, K. Khistyaev, J. Kim, J. Kim, R. A. King, P. Klunzinger, D. Kosenkov, T. Kowalczyk, C. M. Krauter, K. U. Lao, A. Laurent, K. V. Lawler, S. V. Levchenko, C. Y. Lin, F. Liu, E. Livshits, R. C. Lochan, A. Luenser, P. Manohar, S. F. Manzer, S.-P. Mao, N. Mardirossian, A. V. Marenich, S. A. Maurer, N. J. Mayhall, C. M. Oana, R. Olivares-Amaya, D. P. O’Neill, J. A. Parkhill, T. M. Perrine, R. Peverati, P. A. Pieniazek, A. Prociuk, D. R. Rehn, E. Rosta, N. J. Russ, N. Sergueev, S. M. Sharada, S. Sharmaa, D. W. Small, A. Sodt, T. Stein, D. Stück, Y.-C. Su, A. J. W. Thom, T. Tsuchimochi, L. Vogt, O. Vydrov, T. Wang, M. A. Watson, J. Wenzel, A. White, C. F. Williams, V. Vanovschi, S. Yeganeh, S. R. Yost, Z.-Q. You, I. Y. Zhang, X. Zhang, Y. Zhou, B. R. Brooks, G. K. L. Chan, D. M. Chipman, C. J. Cramer, W. A. Goddard III, M. S. Gordon, W. J. Hehre, A. Klamt, H. F. Schaefer III, M. W. Schmidt, C. D. Sherrill, D. G. Truhlar, A. Warshel, X. Xua, A. Aspuru-Guzik, R. Baer, A. T. Bell, N. A. Besley, J.-D. Chai, A. Dreuw, B. D. Dunietz, T. R. Furlani, S. R. Gwaltney, C.-P. Hsu, Y. Jung, J. Kong, D. S. Lambrecht, W. Liang, C. Ochsenfeld, V. A. Rassolov, L. V. Slipchenko, J. E. Subotnik, T. Van Voorhis, J. M. Herbert, A. I. Krylov, P. M. W. Gill, and M. Head-Gordon. Advances in molecular quantum chemistry contained in the Q-Chem 4 program package.,\textit{ Mol.Phys.} \textbf{113}, 184 (2015)

\bibitem{QChem} Shao, Y., \textit{et al.}, Advances in molecular quantum chemistry contained in the Q-Chem 4 program package.,\textit{ Mol.Phys.} \textbf{113}, 184 (2015)



\bibitem{kresse1996efficient} Kresse, G. and Furthm{\"u}ller, J. Efficient iterative schemes for ab initio total-energy calculations using a plane-wave basis set, \textit{Phys. Rev. B} \textbf{54}, 11169 (1996)

\bibitem{kresse1999ultrasoft} Kresse, G. and Joubert, D., From ultrasoft pseudopotentials to the projector augmented-wave method, \textit{Phys. Rev. B} \textbf{59}, 1758 (1999)

\bibitem{tang2009grid} Tang, W., Sanville, E., and Henkelman, G. A grid-based Bader analysis algorithm without lattice bias, \textit{J. Phys.:Cond. Matt.} \textbf{21}, 084204 (2009)

\bibitem{Stukowski2009} Stukowski, A. Visualization and analysis of atomistic simulation data with OVITO–the Open Visualization Tool, \textit{Modelling Simul. Mater. Sci. Eng.} \textbf{18}, 015012 (2010)

\bibitem{Barca2018} Barca, G. M. J., Gilbert, A. T. B., and Gill, P. M. W. Simple Models for Difficult Electronic Excitations. \textit{J. Chem.
Theory Comput.} \textbf{14}, 1501 (2018).

%\bibitem{Anderson_1961}
%Anderson, P. W., Localized Magnetic States in Metals. \textit{Phys. %ev.} \textbf{124}, 1 (1961)

%\bibitem{Izyumov_1965}
%Izyumov, Ya. A. and Medvedev, M.V., Impurity atoms in a %ferromagnetic crystal. \textit{Soviet Physics Journal of Experimental and Theoretical Physics} \textbf{21}, 2 (1965)

%\bibitem{Newns_1969}
%Newns, D. M., Self-Consistent Model of Hydrogen Chemisorption. \textit{Phys. Rev.} \textbf{178}, 3 (1969)

%\bibitem{Popov_2021}
%Popov, I. V., Kushnir, T.S. and Tchougr{\'{e}}eff, A.L., Local perturbations of periodic systems. Chemisorption and point defects by GoGreenGo. \textit{J. Comput. Chem.} \textbf{42}, 32 (2021)

%\bibitem{Ghan_2023}
%Ghan, S., Diesen, E., Kunkel, C., Reuter, K., and Oberhofer, H., %Interpreting ultrafast electron transfer on surfaces with a %converged first-principles Newns–Anderson chemisorption function. %\textit{J. Chem. Phys.} \textbf{158}, 23 (2023)

%TC:endignore
\end{thebibliography}
\end{document}

\title[Article Title]{Timing the Escape of a Caged Electron: Supplementary Information}

\author[1]{\fnm{Connor} \sur{Fields}}

\author[2]{\fnm{Aleksandra} \sur{Foerster}}

\author[2]{\fnm{Sadegh} \sur{Ghaderzadeh}}

\author[2]{\fnm {Ilya} \sur{Popov}}

\author[2]{\fnm{Bang} \sur{Huynh}}

\author[1]{\fnm{Filipe} \sur{Junqueira}}

\author[1]{\fnm{Tyler} \sur{James}}

\author[1]{\fnm{Sofia} \sur{Alonso Perez}}

\author[2,3]{\fnm{David A} \sur{Duncan}}

\author[3]{\fnm{Tien-Lin} \sur{Lee}}

\author[1]{\fnm{Yitao} \sur{Wang}}

\author[4]{\fnm{Sally} \sur{Bloodworth}}

\author[4]{\fnm{Gabriela} \sur{Hoffman}}

\author[4]{\fnm{Mark} \sur{Walkey}}

\author[4]{\fnm{Richard J} \sur{Whitby}}

\author[4]{\fnm{Malcolm H} \sur{Levitt}}

\author[1]{\fnm{Brian} \sur{Kiraly}}

\author[1]{\fnm{James N} \sur{O'Shea}}

\author[2]{\fnm{Elena} \sur{Besley}}

\author[1]{\fnm{Philip} \sur{Moriarty}}

\affil[1]{\orgdiv{School of Physics \& Astronomy}, \orgname{University of Nottingham}, \orgaddress{\street{University Park}, \city{Nottingham}, \postcode{NG7 2RD}, \country{UK}}}

\affil[2]{\orgdiv{School of Chemistry}, \orgname{University of Nottingham},\orgaddress{\street{University Park}, \city{Nottingham}, \postcode{NG7 2RD}, \country{UK}}}

\affil[3]{\orgdiv{Diamond Light Source}, \orgname{Harwell Science \& Innovation Campus}, \orgaddress{\city{Didcot}, \postcode{OX11 0QX}, \country{UK}}}

\affil[4]{\orgdiv{School of Chemistry}, \orgname{University of Southampton},\orgaddress{ \city{Southampton}, \postcode{SO17 1BJ}, \country{UK}}} 
%TC:ignore
\abstract{This supplementary information file comprises additional data and analysis for the following experimental and theoretical aspects of the \textit{Timing the Escape of a Caged Electron} article: \textbf{(i)}~Circumventing beam damage; \textbf{(ii)}~Valence band photoemission for bulk Ar@C$_{60}$; \textbf{(iii)}~Comparison of Ar 2\textit{p} core-level photoemission and Ar 2\textit{p} $\rightarrow$ 4\textit{s} X-ray absorption spectra; \textbf{(iv)}~Fitting the Auger-Meitner decay spectra;  \textbf{(v)}~Comparison of photoemission lineshapes for bulk and monolayer Ar@C$_{60}$ samples; Doniach-Sunjic asymmetry;  \textbf{(vi)}~Raw on-resonance decay spectra for multilayer and monolayer Ar@C$_{60}$ coverage; \textbf{(vii)}~Population analysis: Ar and C contributions to excited state; \textbf{(viii)}~Orbital symmetry decomposition; \textbf{(ix)}~Comparison with time-dependent DFT; \textbf{(x)}~Relativistic considerations; \textbf{(xi)}~Z+1 approximation and ground state K@C$_{60}$; and \textbf{(xii)}~Ar-Ag(111) separation: Additional DFT calculations.}
\pacs{79.20.Fv,  73.50.-h, 71.20.Tx, 82.30.Fi, 31.15.E} 
\keywords{endofullerene, core hole clock, charge transfer, X-ray standing wave, density functional theory, Auger-Meitner electron spectroscopy}
%TC:endignore

\maketitle

\newpage
\section{Circumventing beam damage}
In previous synchrotron-based work -- both published\cite{Jarvis2021} and unpublished -- on endofullerene samples, we have found that measurements acquired at room temperature and without any adjustment of the undulator output flux (and/or sample position) can result in significant beam damage. X-ray absorption and photoemission peaks would, at best, diminish in intensity on a timescale of minutes. While for some samples -- N$_2$@C$_{60}$, for example -- only a reduction in the intensity of the encapsulate-related core-level peak(s) was observed as a function of beam exposure, for other samples (e.g. H$_2$O@C$_{60}$), the line shape and peak position would also change. We note that DiCamillo \textit{et al.}\cite{DiCamillo1996} have reported similar beam damage observations, i.e. the loss of Ar 2$p$ signal, in their lab-based X-ray photoelectron spectroscopy (XPS) studies of Ar@C$_{60}$ (although Morscher \textit{et al.}\cite{Morscher2010} instead did not observe depletion of argon under either Mg K$\alpha$ or He I radiation.)  \\

\noindent Our approach to minimising beam damage for the Ar@C$_{60}$-related measurements described in the main paper and this supplement involved (a) acquiring spectra at low sample temperatures (a maximum of 180 K), and (b) detuning the undulator so as to reduce the photon flux on the sample by an order of magnitude. Throughout the beamtime experiments we regularly checked for evidence of beam damage by comparing photoemission and X-ray absorption peak intensities. No degradation of signal intensity, or other characteristics such as lineshape, was observed for either soft X-ray (photoemission, X-ray absorption (XAS), resonant Auger/photoemission) or hard X-ray (normal incidence X-ray standing wave (NIXSW)) spectroscopies. 

\section{Valence band photoemission for bulk Ar@C$_{60}$ film}
\begin{figure}[h!]
\centering
\includegraphics[scale=0.3]{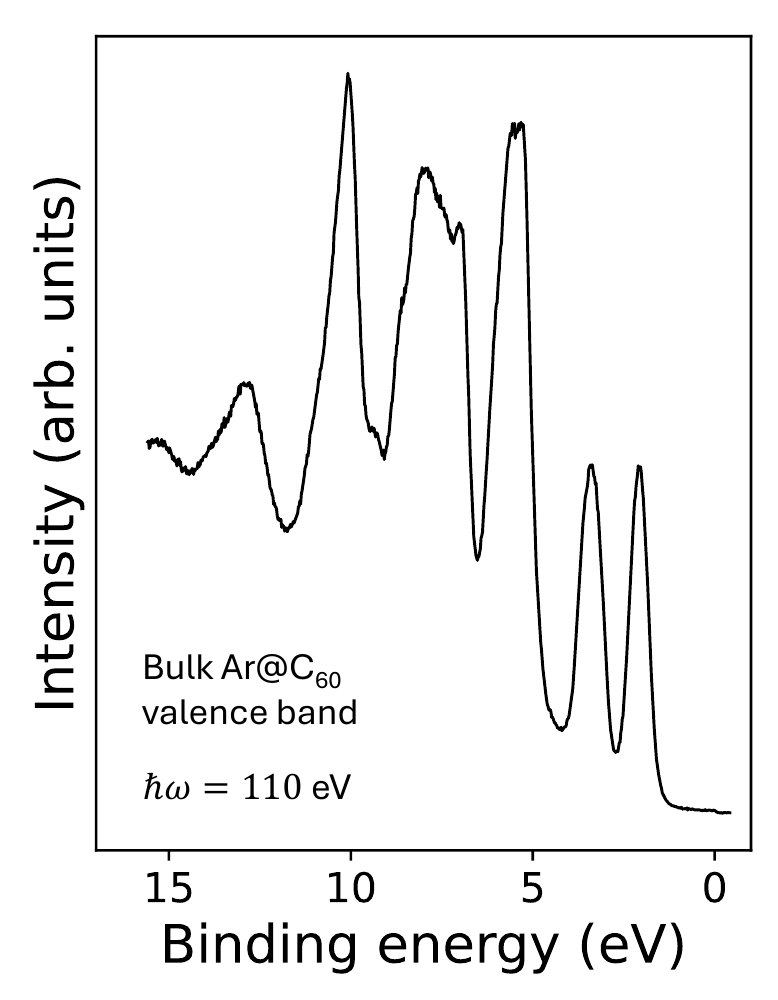}
\caption{Valence band spectrum for a bulk film of Ar@C$_{60}$ acquired with a photon energy of 110 eV.} 
\label{FigS1}
\end{figure}

The valence band spectrum for a thick film of Ar@C$_{60}$ is shown in Fig.\ref{FigS1}. As observed previously by Morscher \textit{et al.}\cite{Morscher2010}, the valence band spectra of empty C$_{60}$ and Ar@C$_{60}$ are indistinguishable down to a binding energy of $\sim$ 10 eV below the Fermi level.

\newpage 
\section{Ionic vs excitonic: Ar 2$p$ XPS-XAS comparison }

\begin{figure}[b!]
\centering
\includegraphics[scale=0.42]{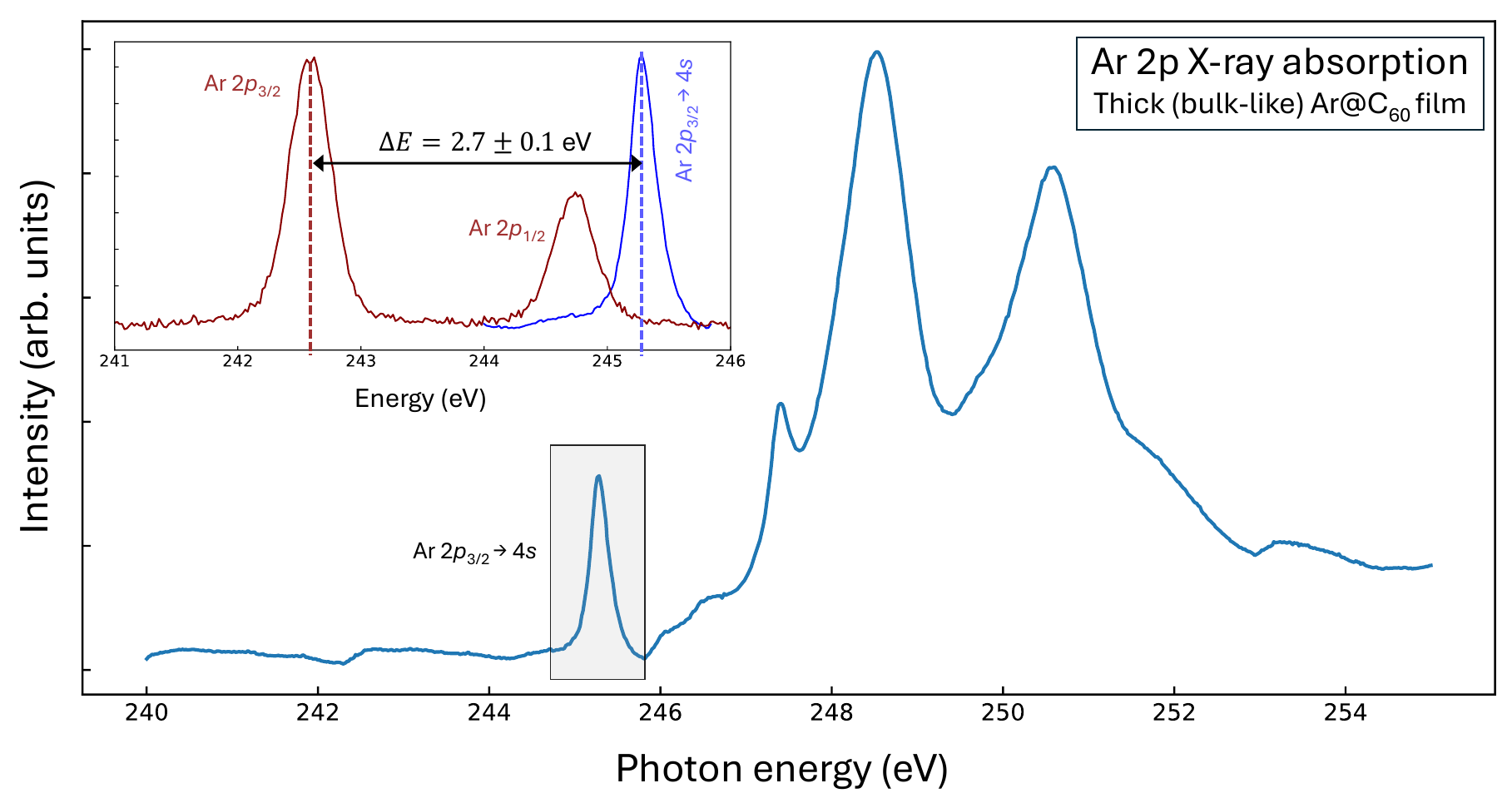}
\caption{\textbf{Ar 2$p$ photoemission and X-ray absorption spectra for the bulk Ar@C$_{60}$ sample.} Wide-energy-range X-ray absorption spectrum for energies close to the Ar 2$p$ (i.e. $L_{2,3}$) absorption edges, measured by recording the sample drain current as a function of photon energy -- a total electron yield measurement. Throughout the work described in the main paper and this supplement, we have focussed on absorption across the Ar $2p_{3/2} \rightarrow 4s$ resonance highlighted by the gray rectangle. See Wurth \textit{et al.}\cite{Wurth1993} for a description of the origin of the other, higher lying resonances in the context of condensed ``bare'' argon multilayers. \textbf{Inset:} Ar 2$p$ core-level photoemission spectrum (\textcolor{red}{in red}) plotted on same energy axis as the Ar $2p \rightarrow 4s$ X-ray absorption spectrum (\textcolor{blue}{in blue}). The energy of the core-level photoemission spectrum is referenced to the Fermi level and the 2$p_{3/2}$-related spectra have been normalised to equivalent heights for ease of comparison.   
}\label{FigS2}
\end{figure}

Fig.\ref{FigS2} shows the Ar 2$p_{3/2} \rightarrow 4s$ X-ray absorption resonance in relation to the Ar 2$p$ ($L_{2,3}$) spectrum acquired over a wider photon energy range. Wurth \textit{et al.}\cite{Wurth1993} have previously identified the origin of the higher lying resonances. In the inset to Fig.\ref{FigS2} we show a comparison of the Ar 2$p$ core-level and $2p_{3/2} \rightarrow 4s$ X-ray absorption spectra, plotted on the same energy scale. (The binding energy (BE) of the core-level spectrum is referenced to the Fermi level.) There is a difference of 2.7 $\pm$ 0.1~eV between the Ar 2$p_{3/2}$ core level BE and the peak energy of the $2p \rightarrow 4s$ X-ray absorption spectrum, arising from the ionic (photoemission) vs neutral (excitonic, X-ray absorption) character of the final state of each process.\\ 

\noindent As highlighted by Martensson \textit{et al.}\cite{Martensson1999} and Sandell \textit{et al.}\cite{Sandell1999,Sandell1997} in the context of argon adsorption on graphite, the observation that the Ar 2$p_{3/2}$ binding energy is significantly \textit{lower} than that of the X-ray absorption peak is already a clear indication that the ionic state is considerably more energetically favourable than the excitonic, charge neutral state -- this is ultimately the driving force for transfer of the 4$s$ electron to the environment. We note that the difference of 2.7 $\pm$ 0.1 eV is rather higher than the value of 2.1 eV observed for argon on graphite\cite{Sandell1999}, but identical to the 2.72 eV value reported by Lizzit \textit{et al.}\cite{Lizzit2013} for argon on a weakly coupled graphene monolayer on a ruthenium substrate. (Oxygen intercalation was used to reduce the coupling to the metal surface.) \\

\noindent As the photoemission binding energy, $E_{XPS}$, is referenced to the Fermi level, $E_F$, the difference between this and the X-ray absorption energy, i.e. $\Delta E = E_{XAS}-E_{XPS}$ is generally taken to represent the energy of the excitonic state with respect to $E_F$. This would in turn place the Ar 4$s$ state at 4.90 $\pm$~0.15 eV above the HOMO level (see Fig. 1(d) of the main paper for a valence band spectrum.) Lizzit \textit{et al.}\cite{Lizzit2013} probe the alignment of the Ar 4$s$ level with the empty density of states of graphene in this way; we adopt a similar approach in the main paper to ascertain the extent to which the core-excited Ar 4$s$ state might align energetically with the $s$-SAMO (superatomic orbital) level. We note, however, that while the agreement with the 4.9 eV separation of the s-SAMO level from the HOMO in the calculations of Zhao \textit{et al.}\cite{Zhao2009} is certainly noteworthy, the alignment of energy levels calculated from our experimental data in this way ignores the possible influence of the Ar 2$p$ core hole on the HOMO and s-SAMO states of the fullerene cage. 

\section{Fitting Auger-Meitner decay spectra}
Although the positions and relative intensities of the primary, easily identifiable spectator and normal Auger-Meitner features (at binding energies of $\sim$ 26.8~eV, 28.5~eV, 30.8~eV (on resonance), 33.1~eV (on resonance), 31.7~eV, and 32.1~eV) in the map and spectra shown in Fig. 2 of the main paper are in line with many previous studies [see, for example, refs 29, 31, 32, 33, 34, and 36 in the main paper], there is one significant difference. We found a particular issue with fitting the peaks between binding energies of $\sim$ 29 eV and 30 eV -- it was not possible to robustly track the normal Auger- Meitner feature that appears at an on-resonance binding energy of $\sim$ 30 eV throughout the entire 1 eV width of the resonance region while keeping the intensity ratio for all normal Auger-Meitner peaks constant, not least because this feature overlaps heavily with the spectator contribution. This necessarily leads to a higher level of uncertainty in the division of spectral intensity into normal Auger-Meitner and spectator contributions.\\

\begin{figure}[h!]
\centering
\includegraphics[scale=0.55]{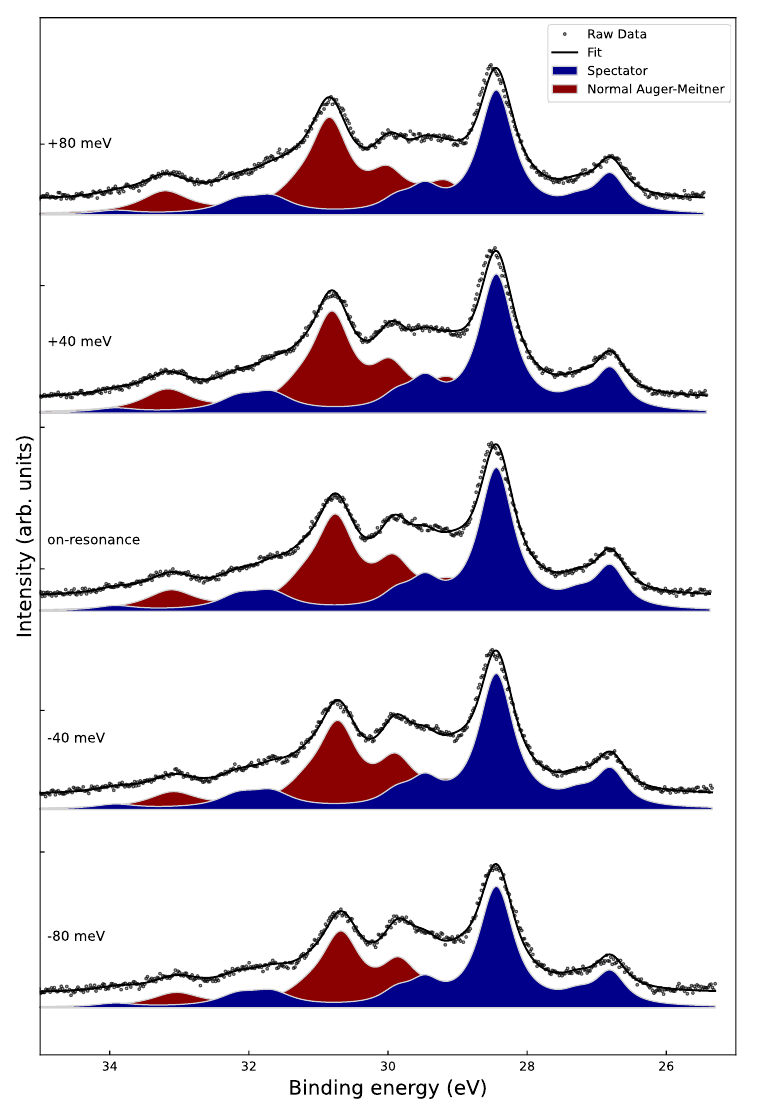}
\caption{\textbf{Fitted decay spectra.} A set of decay spectra, with accompanying fits, acquired with photon energies either side of the on-resonance condition ($\hbar \omega=245.3$ eV).}  
\label{FigS3}
\end{figure}

\noindent Fig.\ref{FigS3} shows a comparison of fits of the decay spectra either side of the resonance condition ($\hbar \omega = 245.3$ eV). Our strategy in fitting these (and all other decay spectra) was to fully constrain all peak positions and peak widths, i.e. to allow no freedom in those parameters as a function of photon energy. In other words, the fitted binding energy of the spectator peaks across the resonance does not vary, and the normal Auger-Meitner peaks rigidly change their binding energy in line with the variation in photon energy (i.e. a change of 20 meV from spectrum to spectrum). In addition, the relative intensities of the spectator peaks (including the shake-up components) are fully constrained. However, in order to provide satisfactory fits, we allowed a small variation in the relative intensities of the Auger-Meitner peaks with on-resonance binding energies in the 29 - 31 eV range.\\

\noindent Attempts to account for this spectral intensity by fitting an additional, shifted normal Auger-Meitner contribution arising from a small residual contribution from the first Ar@C$_{60}$ monolayer directly bonded to the Ag(111) substrate (see Section 6 below) were not successful. Our motivation in attempting to introduce this contribution was informed by our sample preparation strategy: depositing the argon endofullerene onto a cold substrate significantly limits molecular mobility and it is very unlikely that the film grows in a simple layer-by-layer mode. Indeed, the slow rate of decay of the Ag 3$d$ and Ag(111) valence band photoemission intensity as a function of deposition time would point to significant islanding, with regions of the first chemisorbed monolayer remaining exposed up to high molecular coverages. (There is an appreciable Ehrlich–Schwoebel barrier ($\sim$ 120 meV) for C$_{60}$ diffusion\cite{Bommel2014}.)

\noindent However, a number of observations led us to discount the possibility of a residual contribution to the multilayer/bulk spectra arising from the monolayer: (i) as discussed in the main paper and Section 6 below, there is a significant, 1.25 eV, shift in the kinetic energies of the normal Auger-Meitner peaks for the endofullerene monolayer as compared to the bulk film. The normal Auger-Meitner intensity in the 29 -- 30 eV binding energy range is not in line with this shift; (ii) as discussed in the following section, there is a very clear binding energy difference in the C 1$s$ and Ar 2$p$ core-level emission for the monolayer vs the bulk film that would produce a strong shoulder in the spectra should the monolayer make more than a negligible contribution. We did not observe a shoulder of this type in the multilayer specta; and (iii) the monolayer Auger-Meitner emission has a slightly higher on-resonance energy than that for the bulk film (245.4 eV vs 245.3 eV), out of line with the behaviour of the spectral intensity in the 29-30 eV binding energy (on resonance) range. \\

\begin{figure}[h!]
\centering
\includegraphics[scale=0.3]{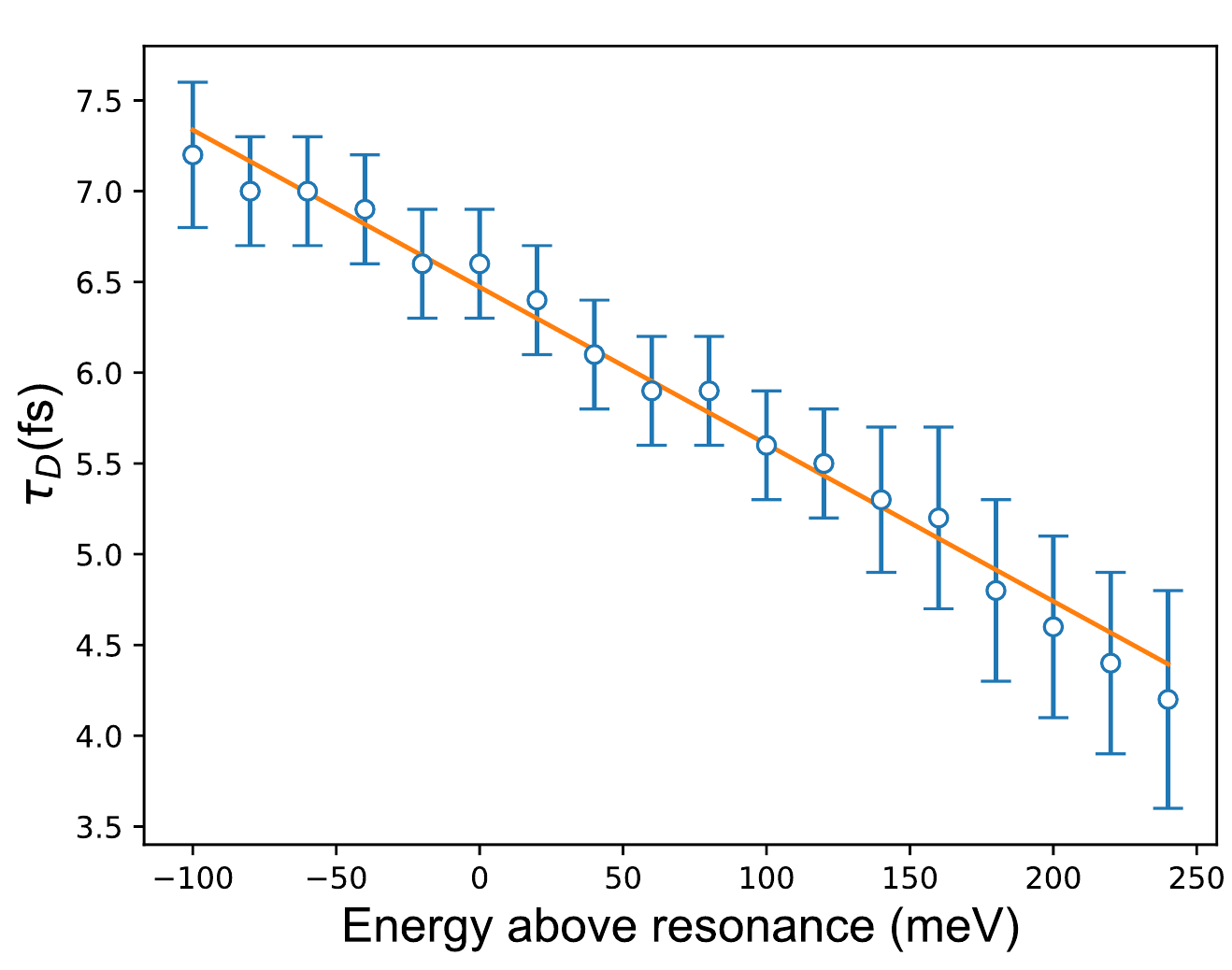}
\caption{\textbf{Variation of delocalisation time with photon energy.} There is a monotonic decrease in $\tau_D$ with increasing photon energy in this range. We have fitted a linear dependence. Other functional forms could in principle also be used to fit the data, including an exponential decay in line with the dependence of tunnelling rate on effective barrier height, but are difficult to justify due to the magnitude of the uncertainties.}    
\label{FigS4}
\end{figure}

\noindent There is one final key point on the subject of fitting the decay spectra that we would like to raise. Given the large number of constrained parameters, uncertainties that are returned via the standard approach, i.e. diagonalisation of the covariance matrix produced via the non-linear least squares fitting routine (which we implemented using the Python LMFIT package), are often significantly underestimated (and, moreover, assume normally distributed errors.) In order to provide a more robust estimate of the uncertainties, we therefore allowed the parameters to vary by a small amount about their constrained values and employed a Monte Carlo/bootstrapping approach to explore the fitness landscape. For decay spectra close to resonance (i.e. those excited with photon energies within $\pm$ 100 meV of the on-resonance value of 245.3 eV), we estimate that the value of $\tau_D$ can be determined to an uncertainty of approximately 5\%. Outside of this photon energy range, the signal-to-noise ratio is appreciably larger and the estimated uncertainties consequently significantly larger (up to 10\%). \\

\noindent Fig.\ref{FigS4} shows the variation in $\tau_D$ as a function of photon energy around the resonance condition. There is a monotonic decrease in the delocalization time across the resonance, which we have fitted with a simple linear dependence. The magnitude of the uncertainties does not justify a more sophisticated/higher order fitting function. \\   

\noindent Although better fits -- i.e., lower $\chi ^2$ values and, nominally, lower uncertainties -- could be obtained by allowing small (i.e. few percent) variations in peak positions, widths, and intensity ratios, we rejected this fitting strategy. Given the exceptionally large parameter space, we preferred instead, and as noted above, to heavily constrain the fits on the basis of the relevant physics/physical chemistry of the system. (As von Neumann famously said, ``\textit{With four parameters I can fit an elephant, and with five I can make him wiggle his trunk.}''\cite{Mayer2010}.)  

\section{Comparison of photoemission spectra for bulk and monolayer Ar@C$_{60}$}

\begin{figure}[b!]
\centering
\includegraphics[scale=0.4]{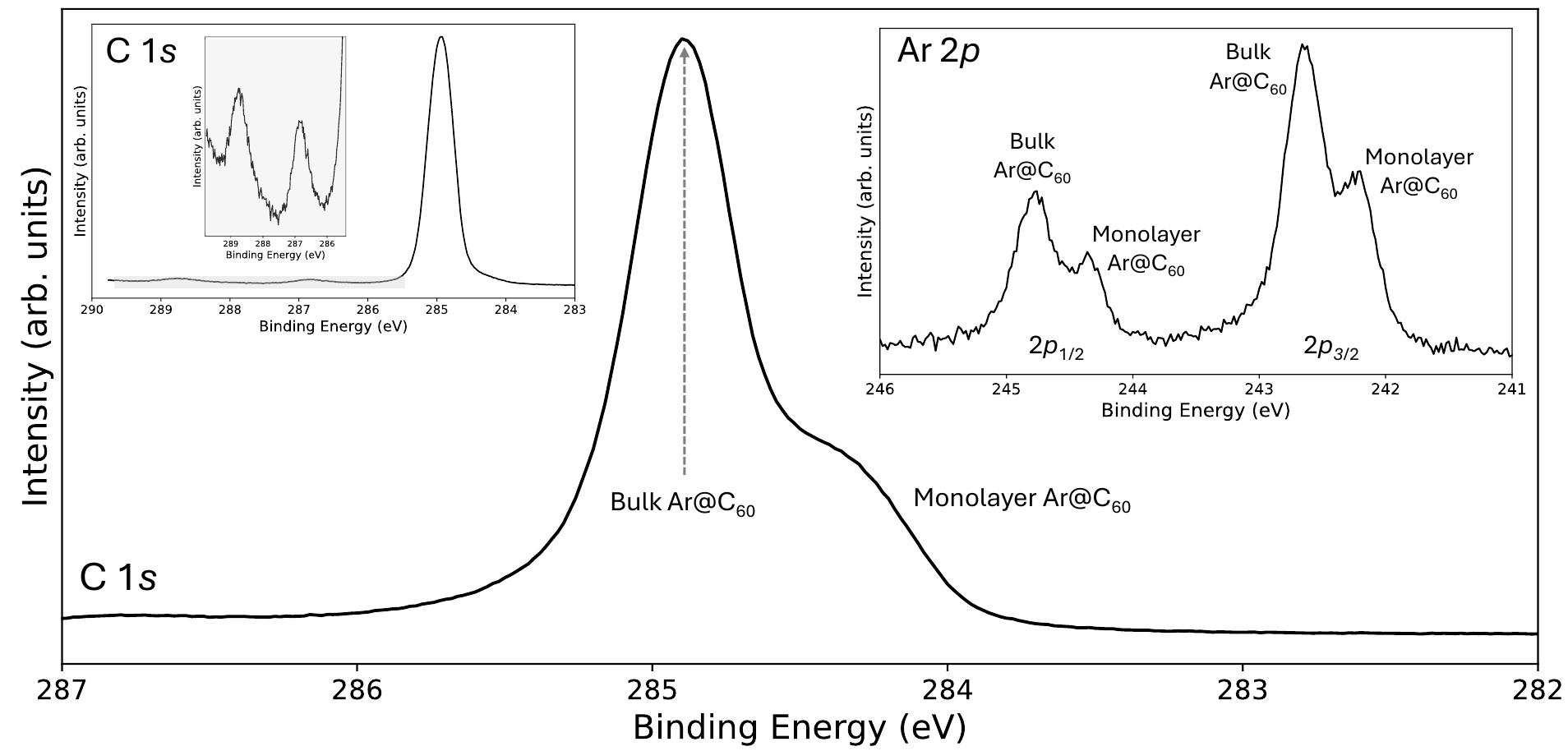}
\caption{\textbf{Comparison of photoemission spectra for monolayer and multilayer coverages.} The main figure is a C 1$s$ photoemission spectrum (acquired with a photon energy of 400 eV) of a multilayer coverage of Ar@C$_{60}$ on Ag(111) that clearly shows the significant (400~meV) difference in core-level binding energy associated with endofullerene molecules in the first monolayer and in higher layers. \textbf{Inset to left:} C 1$s$ spectrum for a bulk-like Ar@C$_{60}$ film sufficiently thick to ``quench'' Ag 3$d$ photoemission intensity from the Ag(111) substrate on which it was grown. The shake-up features are magnified in the inset. This spectrum is indistinguishable from that of empty C$_{60}$. \textbf{Inset to right:} Ar 2$p$ photoemission spectrum ($\hbar\omega =$~400 eV) for the same multilayer sample as shown in the main figure. Again, there is a 400~meV shift towards lower binding energy for photoemission from molecules in the first monolayer, as compared to those in higher layers.     
}\label{FigS5}
\end{figure}

There is a substantial shift in both C 1$s$ and Ar 2$p$ core-level binding energy (CLBE) for Ar@C$_{60}$ molecules that are bound in the first monolayer to the Ag(111) surface as compared to the CLBE for endofullerenes in higher (physisorbed) layers. This BE shift has also previously been observed for empty C$_{60}$ on Ag(111)\cite{Pedio1999} (see below for a detailed comparison). The main panel in Fig.\ref{FigS5} highlights the BE shift (but obscures the difference in C1$s$ lineshape for monolayer vs bulk Ar@C$_{60}$ samples -- we'll return to this soon.) Argon 2$p$ photoemission spectra (inset to right in Fig.\ref{FigS5}) show the same 400~meV binding energy difference for monolayer versus bulk Ar@C$_{60}$ samples as observed for the C 1$s$ data. \\

\noindent A bulk film of Ar@C$_{60}$, sufficiently thick so that minimal Ag 3$d$ emission from the Ag(111) susbtrate is observed, yields the C 1$s$ spectrum shown in the upper left inset to Fig.\ref{FigS5}. This, and the associated shake-up spectrum (also shown in Fig.\ref{FigS5}), are identical to the corresponding spectra for empty C$_{60}$, both in terms of line shape and binding energy. 

\subsection{Lineshape analysis: Doniach-Sunjic asymmetry}
\begin{figure}[t!]
\centering
\includegraphics[scale=0.4]{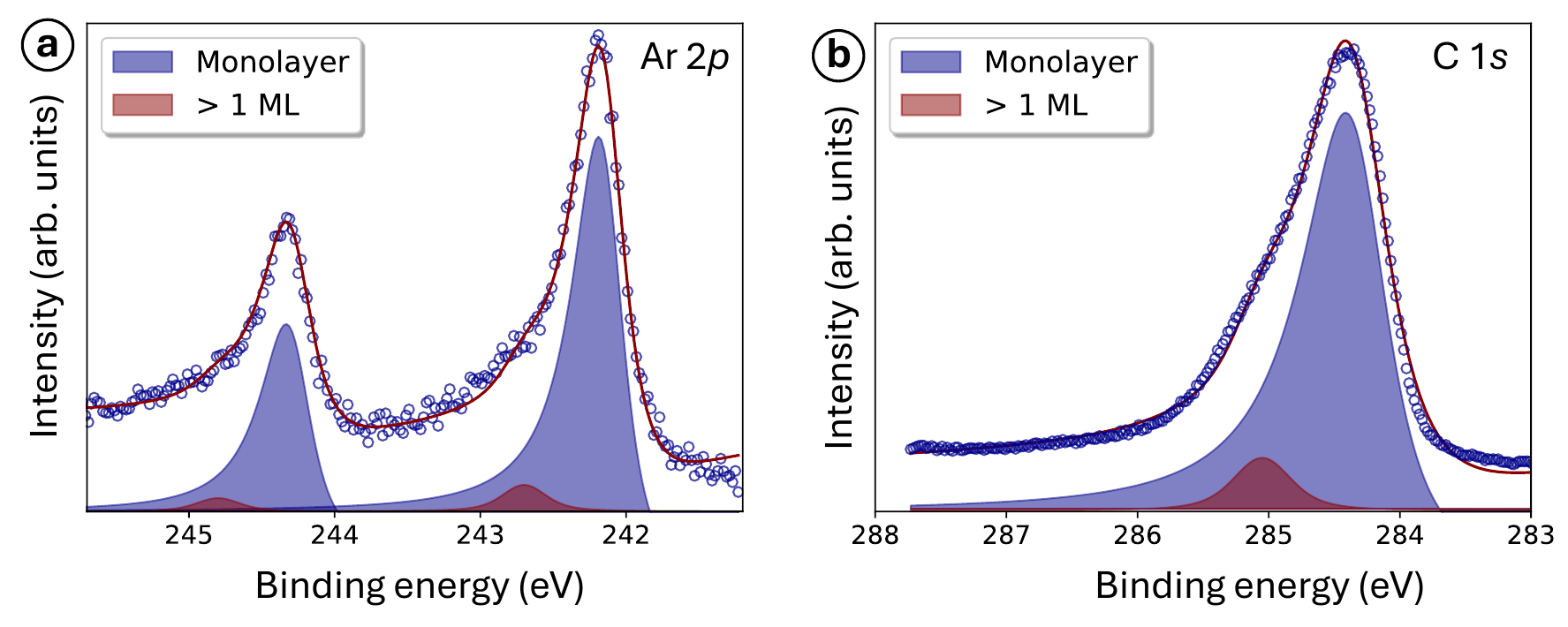}
\caption{\textbf{Doniach-Sunjic asymmetry in monolayer photoemission lineshapes.} The \textbf{(a)} Ar~2$p$ and \textbf{(b)} C~1$s$ photoemission lineshapes ($\hbar\omega=400$ eV) for the Ar@C$_{60}$ monolayer show a pronounced asymmetry to higher binding energy that arises from the response of the metallic conduction electron gas to the creation of the core hole\cite{DS1970,ArBubbles}. We have fitted the asymmetric lineshape in each case to a Doniach-Sunjic function\cite{DS1970}, but also have to account for an additional asymmetry due to a small amount of additional adsorbed endofullerene above the first monolayer (shaded in \textcolor{purple}{purple} in the fits above). Each peak compromises two components: one for the monolayer (fitted with an asymmetric DS profile), and a second, much lower intensity, contribution at higher binding energy arising from physisorbed endofullerene on top of the chemisorbed monolayer. The contributions from the physiorbed molecules are each fitted with a symmetric Voigt lineshape. The asymmetry parameters, $\alpha$, for the C 1$s$ and Ar 2$p$ spectra, 0.243 $\pm$ 0.007 and 0.28 $\pm$ 0.02, respectively, are very similar. (The error bars are determined from the covariance matrix of the fit.) 
}\label{FigS6}
\end{figure}

Although the position of the argon atom inside the endofullerene is unaffected by the significant level of charge transfer from the Ag(111) surface into the cage at the molecule-surface interface (see main paper), the Ar~2$p$ photoemission spectrum for a monolayer coverage of Ar@C$_{60}$ on Ag(111) nonetheless shows the signature asymmetry, i.e. the Doniach-Sunjic (D-S) lineshape\cite{DS1970}, arising from excitation of the conduction electron density of the underlying metal (Fig.\ref{FigS6}).  As also observed by both Pedio \textit{et al.}\cite{Pedio1999} and Gibson \textit{et al.}\cite{Gibson2017} for 1ML of empty C$_{60}$ on Ag(111), the C 1$s$ core-level spectrum is similarly highly asymmetric. (See below.) In other words, while the encapsulated argon atom is unaffected by endofullerene adsorption from the perspective of its intracage position, it is nonetheless highly sensitive to the surrounding electrostatic and electrodynamic environment\cite{MeierNote}.\\

\begin{figure}[h!]
\centering
\includegraphics[scale=0.3]{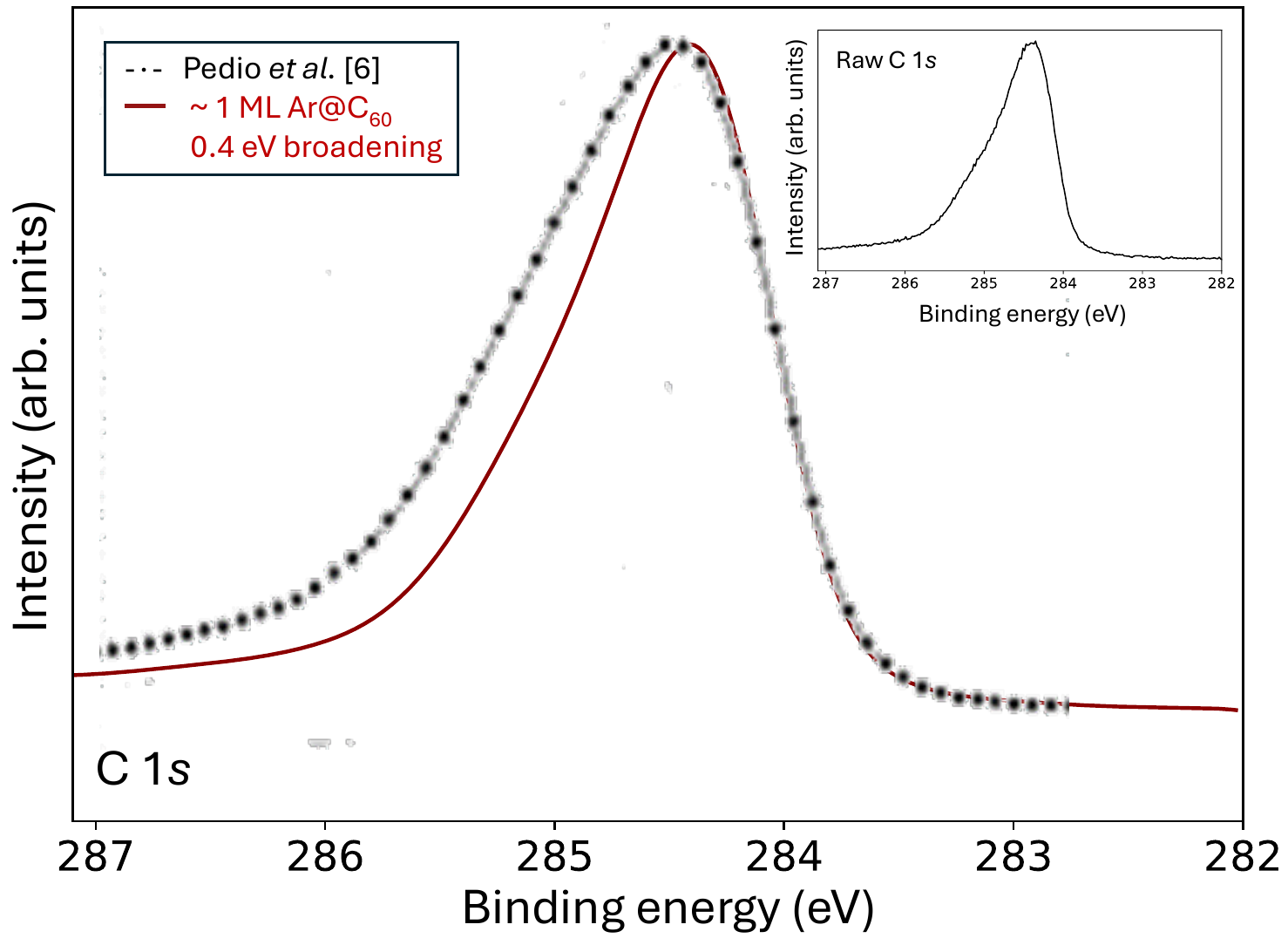}
\caption{\textbf{Comparison of C 1$s$ spectrum for monolayer C$_{60}$ and Ar@C$_{60}$ on Ag(111).\\} \textcolor{red}{Red line}: C 1$s$ spectrum for $\sim$ 1 ML Ar@C$_{60}$ on Ag(111) artificially broadened by convolution with a Gaussian of 0.4 eV FWHM to match experimental resolution of Pedio \textit{et al.}\cite{Pedio1999}, who reported C~1$s$ X-ray photoelectron spectroscopy (XPS) spectra for a variety of C$_{60}$-on-metal systems, including C$_{60}$ on Ag(111). Pedio \textit{et al.}'s spectrum is shown as the filled-circles-with-line curve and has been digitally cut from Fig. 4 of their paper\cite{Pedio1999}, scaled appropriately, aligned, and superimposed on the broadened spectrum for Ar@C$_{60}$. (The original, unbroadened C 1$s$ spectrum for $\sim$ 1 monolayer of Ar@C$_{60}$ on Ag(111) is shown in the inset.) 
}\label{FigS7}
\end{figure}

\noindent Notably, the asymmetry parameters for the D-S lineshapes are very similar -- 0.243 $\pm$ 0.007 (C 1$s$) and 0.28 $\pm$ 0.01 (Ar 2$p$), indicative of a common origin. (The fitting is complicated somewhat by the presence of a small amount of endofullerene above the first monolayer due to the sample preparation process, where a small ``overshoot'' in coverage is difficult to avoid. However, the relatively large BE shift of 0.4 ($\pm 0.1$) eV between the Ar 2$p$ core-level signal for the first endofullerene layer and for subsequent layers facilitates easy identification of this contribution to the photoemission spectra. The endofullerene molecules above the first layer also make a characteristic contribution to the low energy electron diffraction (LEED) pattern, forming a weak multi-domain pattern that is distinct from the $(2\sqrt{3} \times 2\sqrt{3}$)R30$^{\circ}$ superlattice.)  \\

\noindent In order to better compare our Ar@C$_{60}$ photoemission data with previously published (and extensive) work on the empty C$_{60}$-on-Ag(111) system\cite{Pedio1999}, we have artificially broadened (via numerical convolution) our C 1$s$ spectrum for the close-to-1-ML Ar@C$_{60}$ sample with a Gaussian having a full width at half maximum of 0.4 eV. This is to match the experimental resolution reported by Pedio \textit{et al.}\cite{Pedio1999} for their C 1$s$ X-ray photoelectron spectroscopy (XPS) measurements. This comparison is shown in Fig.\ref{FigS7}. Note that, despite the presence of a small amount of additional enodfullerene material above the first chemisrobed monolayer, the width of the C 1$s$ peak for Ar@C$_{60}$/Ag(111) is significantly narrower than that for empty C$_{60}$ on Ag(111). We attribute this to the difference in sample preparation conditions. Pedio \textit{et al.} deposited fullerene molecules onto a sample that was at room temperature. The Ar@C$_{60}$ monolayer was instead formed via deposition onto a Ag(111) crystal held at 180 K, circumventing ``nanopitting'' at the fullerene-Ag(111) interface\cite{Diehl2009} and leading to both a weaker and more homogeneous cage-surface interaction.

\section{Decay spectra for multilayer and monolayer Ar@C$_{60}$ coverage} 
\begin{figure}[h!]
\centering
\includegraphics[scale=0.3]{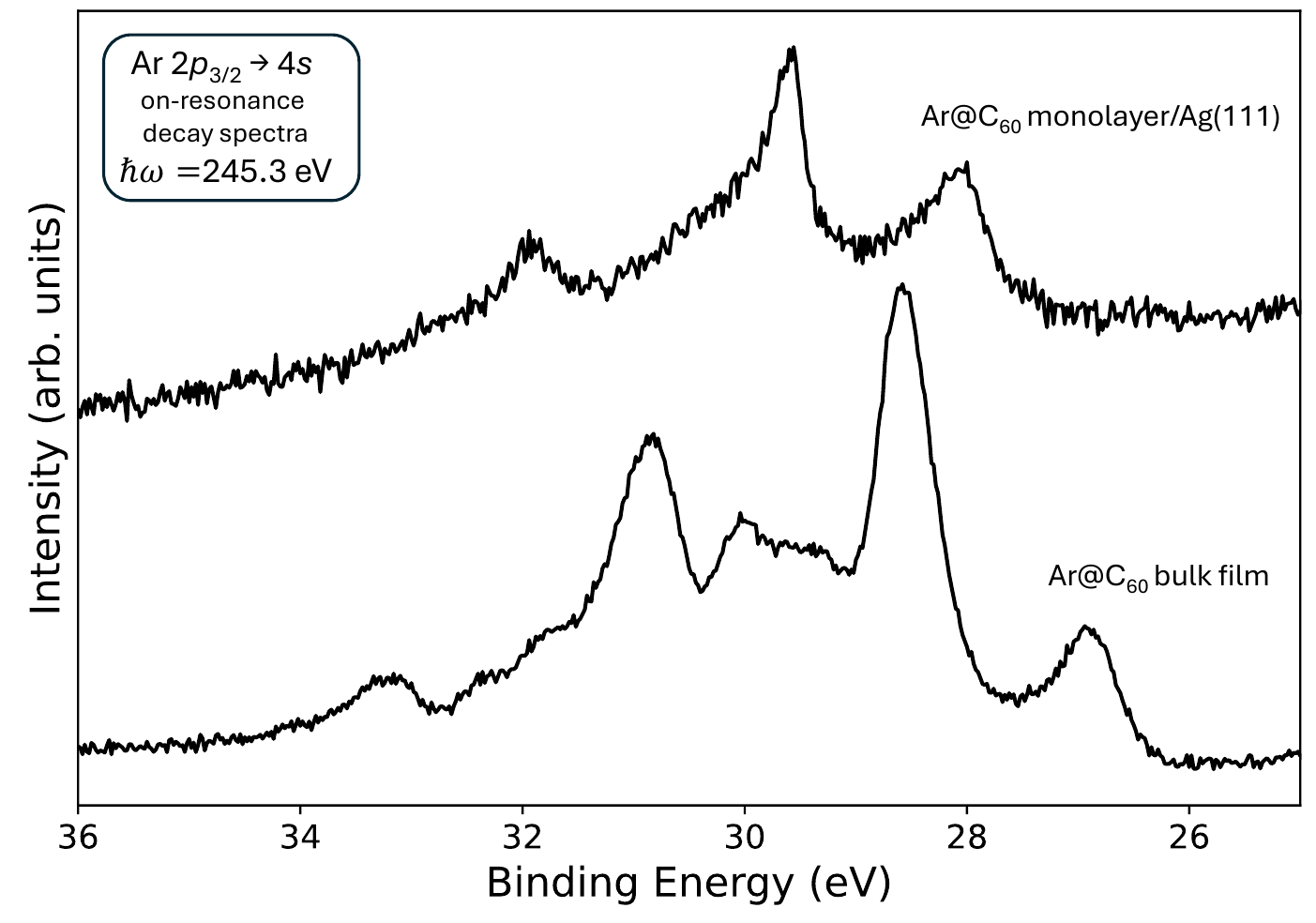}
\caption{\textbf{Decay spectra for multilayer vs monolayer.} Fig. 5 of the main paper shows a comparison between the Auger/spectator decay spectra for multilayer and monolayer coverages, following removal of a linear background and a shift of binding energy of the latter to bring it into alignment with the multilayer spectrum. Here we show the raw spectra before adjustment (other than removal of a constant background for display purposes). 
}\label{FigS8}
\end{figure}

In Fig. 5 of the main paper we show a comparison of the Auger spectrum for an Ar@C$_{60}$ multilayer and monolayer for which we have aligned the corresponding Auger peaks and also subtracted a linear background from the spectrum for the Ar@C$_{60}$ monolayer. For completeness, we show here (Fig.\ref{FigS8}) the raw data before adjustment.

\section{Population analysis: Argon and carbon contributions to the 4$s$ excited state}

We found that the quantification of argon and carbon contributions to the excited state is less straightforward than one might have anticipated (or hoped for). Essentially, we tackle the notion of ``mixing of argon and fullerene orbitals'’ directly by decomposing each orbital $\psi(\mathbf{r})$ into contributions from the argon atom and the carbon atoms in the fullerene cage. There are, however, many decomposition schemes, with each giving different decompositions that have varying degrees of physicality.\\

The most popular decomposition scheme is that of Mulliken\cite{Mulliken1955}, which simply computes the trace of the density matrix block corresponding to basis functions centred on the atom of interest while ignoring all off-diagonal blocks (see also Section 3.9.3 of the \href{http://sobereva.com/multiwfn/Multiwfn_manual.html}{Multiwfn manual}). Even though this scheme gives reasonable results most of the time, it is susceptible to producing negative decomposition values, which are unphysical. The Ar-contributions to the relevant molecular orbitals in the ground and excited calculations as calculated using Mulliken's decomposition scheme are shown in the Excel sheet, \texttt{popana.xlsx}, uploaded as part of the \href{ http://doi.org/10.17639/nott.7457}{data repository}.\\

\noindent It can be seen that, for all occupied (restricted) molecular orbitals in the ground state, and for all occupied $\alpha$ molecular orbitals except the distorted $4s$ one in the excited state, Mulliken's Ar-contributions are sensible. However, for the unoccupied $4s$ molecular orbital in the ground state and the occupied distorted $4s$ in the excited state, Mulliken's Ar-contributions are strongly negative, which means we cannot use this scheme to say anything meaningful about the molecular orbitals of interest to us.\\
  
\noindent We must therefore seek a more sensible alternative decomposition scheme. The main issue that these schemes must deal with is the balanced treatment of the cross terms that are ignored in Mulliken's scheme. One attempt is that by Stout and Politzer\cite{Stout1968} (see also Section 3.9.6 of the \href{http://sobereva.com/multiwfn/Multiwfn_manual.html}{Multiwfn manual}), but this turns out to be even worse than Mulliken's, as seen in the even more negative Ar-contributions in both the ground and excited $4s$ molecular orbitals.\\

\noindent Other attempts include the use of Becke's weighting function\cite{Becke1988} (see also Section 3.9.8 of the \href{http://sobereva.com/multiwfn/Multiwfn_manual.html}{Multiwfn manual}) and Hirshfeld's atomic charge\cite{Hirshfeld1977} (see also Section 3.9.1 of the \href{http://sobereva.com/multiwfn/Multiwfn_manual.html}{Multiwfn manual}). However, both of these methods rely on a numerical real-space integration scheme for the electron density and give very small Ar-contributions for both the ground and excited $4s$ molecular orbitals, which is clearly inconsistent with the fact that the coefficients for these molecular orbitals contain relatively much larger values for the basis functions centred on the Ar atom compared to those for the basis functions centred on the C atoms of the cage.\\

\noindent Fortunately, there exists an alternative by Ros and Schuit\cite{Ros1966}, \textit{C-squared population analysis (SCPA)}, that makes use of the coefficients directly (rather than via numerical real-space integrations) and that also ensures that the atom contributions are never negative (see also Section 3.9.5 of the \href{http://sobereva.com/multiwfn/Multiwfn_manual.html}{Multiwfn manual}). The Ar-contributions to the ground and excited $4s$ molecular orbitals in the SCPA scheme now appear sensible. The ground unoccupied $4s$ molecular orbital has $\sim$ 92.17\% contribution from the argon atom, which means that there is only $\sim$ 7.83\% of mixing with all the carbon atoms on the cage. On the other hand, the excited occupied and distorted $4s$ molecular orbital only has $\sim$13.38 \% contribution from the argon atom, indicating that there is now significant mixing with the cage. 
\vspace{-3mm}
\section{Orbital symmetry considerations}
In the main paper, we note that we used QSYM$^2$\cite{Huynh2024} to extract the symmetry components of the excited state. The following approach was adopted when assessing the symmetry decompositions.\\

\noindent By applying the projection operator
        \begin{equation}
          \hat{\mathcal{P}}^{\Gamma} =
            \frac{d_{\Gamma}}{\lvert \mathcal{G} \rvert}
            \sum_{\hat{g} \in \mathcal{G}}
              \chi^{\Gamma}(\hat{g})^*
              \hat{g}
        \end{equation}
        to a molecular orbital $\psi$, where $\mathcal{G}$ is the symmetry group of the system, $\Gamma$ an irreducible representation of $\mathcal{G}$ with dimensionality $d_{\Gamma}$, and $\chi^{\Gamma}(\hat{g})$ the character of the symmetry operation $\hat{g}$ in $\Gamma$, we can determine the $\Gamma$-composition of $\psi$:
        \begin{equation}
          c_{\Gamma}(\psi)
            = \braket{\psi | \hat{\mathcal{P}}^{\Gamma} | \psi}
            = \frac{d_{\Gamma}}{\lvert \mathcal{G} \rvert}
              \sum_{\hat{g} \in \mathcal{G}}
              \chi^{\Gamma}(\hat{g})^*
              \braket{\psi | \hat{g} | \psi},
        \end{equation}
        where it can be shown that
        \begin{equation}
          \sum_{\Gamma} c_{\Gamma}(\psi) = \braket{\psi | \psi} = 1
        \end{equation}
        if $\psi$ is normalised.
        This procedure can be carried out using QSYM$^{2}$\cite{Huynh2024}.
        Since the calculations for Ar@C$_{60}$ were done with a perfect icosahedral structure, the group $\mathcal{G}$ is the icosahedral group $\mathcal{I}_h$ and the possible irreducible representations are $A_g$, $T_{1g}$, $T_{2g}$, $F_{g}$, $H_g$, $A_u$, $T_{1u}$, $T_{2u}$, $F_{u}$, and $H_u$.
        The above procedure applied to the ground- and excited-state $4s$ molecular orbitals yields the decomposition shown in \ref{tab:decom}.\\
        
        \noindent We know from the $O(3) \supset \mathcal{I}_h$ subduction that the first few even spherical harmonics have the following symmetries in $\mathcal{I}_h$:
        \begin{alignat*}{3}
          &S &\ (l = 0) &\rightarrow A_g,\\
          &D &\ (l = 2) &\rightarrow H_g,\\
          &G &\ (l = 4) &\rightarrow F_g \oplus H_g,\\
          &I &\ (l = 6) &\rightarrow A_g \oplus T_{1g} \oplus F_g \oplus H_g.
        \end{alignat*}
        Hence, with the decompositions shown above, we deduce that the ground-state $4s$ molecular orbital consists approximately of  $\sim$ \% $S$-component, whereas the excited-state distorted $4s$ molecular orbital consists approximately of $\sim$ 76\% $S$-symmetry component, $\sim$ 23\% $D$-symmetry component, and less than 1\% $G$-symmetry component.
        A direct decomposition of each molecular orbital into spherical-harmonic components is possible but is beyond the scope of this work.

\begin{table}[h!]
  \centering
  \caption{%
    Irreducible representation decompositions for the ground- and excited-state $4s$ molecular orbitals in the $\mathcal{I}_h$ structure.
    Both of these molecular orbitals are even with respect to spatial inversion, and so only the \textit{gerade} irreducible representations need to be considered.}
  \begin{tabular}{
    l | r r
  }
    \toprule
    {$\mathcal{I}_h$} & {ground-state $4s$ (\%)} & {excited-state $4s$ (\%)} \\ \midrule
    $A_g$    & 100.00 & 76.06 \\
    $T_{1g}$ &   0.00 &  0.00 \\
    $T_{2g}$ &   0.00 &  0.00 \\
    $F_{g}$  &   0.00 &  0.67 \\
    $H_{g}$  &   0.00 & 23.27 \\ 
    \bottomrule
  \end{tabular}
  \label{tab:decom}
\end{table}

\section{The use of $\Delta SCF$ via the maximum overlap method (MOM); comparison with time-dependent DFT calculations}
We used $\Delta$SCF with high confidence, as numerous studies have showed that $\Delta$SCF can predict excitation energies with accuracy comparable to, and occasionally surpassing, that of time-dependent DFT (TD-DFT) when employing the maximum overlap method (MOM) to optimise the excited states\cite{BourneWorster2021,Bogo2024,Kowalczyk2011}. In Table \ref{tab:tddft}, we compare the calculated excitation energies for the Ar $2p \rightarrow 4s$ transition between MOM and TD-DFT in two different exchange--correlation functionals. It can be seen that while $\Delta$SCF/MOM somewhat overestimates the excitation energies, TD-DFT \textit{severely} underestimates them.\\

\noindent We attribute the discrepancy between the accuracies of $\Delta$SCF/MOM and TD-DFT compared to experimental results to the fact that the orbitals in TD-DFT calculations are unrelaxed. As such, any TD-DFT excited states involving the unrelaxed unoccupied $4s$ orbital of the ground-state DFT calculation would not be able to pick up the qualitative distortion due to the mixing with the C$_{60}$ cage in the excited state as demonstrated by $\Delta$SCF via the MOM. Moreover, using $\Delta$SCF via the MOM allows us to visualise the orbitals involved in the transition, therefore gaining a clearer qualitative understanding of the targeted excited states, which would not be possible with TD-DFT.\\

\begin{table}[h!]
  \centering
  \caption{
    Calculated excitation energies for the Ar $2p \rightarrow 4s$ transition.
    All energy values are in eV and correspond to singlet excited states with $M_S = 0$ and $\langle \hat{S}^2 \rangle \approx 0$.
    This was ensured in $\Delta$SCF via the approximate spin purification equation \cite{Daul1994,Ziegler1977} $E_{\langle \hat{S}^2 \rangle \approx 0} \approx 2 E_{\langle \hat{S}^2 \rangle \approx 1} - E_{\langle \hat{S}^2 \rangle \approx 2}$, where the MOM was used to converge into Kohn--Sham states with $M_S = 0$ and $\langle \hat{S}^2 \rangle \approx 1$ and $\langle \hat{S}^2 \rangle \approx 2$ corresponding to the Ar $2p \rightarrow 4s$ transition. The experimentally measured energy for the Ar $2p_{3/2} \rightarrow 4s$ transition is approximately 245.3 eV (see Figures 2 and S1).
  }
  \begin{tabular}{
    l | r r
  }
    \toprule
                      & $\Delta$SCF/MOM & TDDFT \\ \midrule
    PBE/6-31++G**     & 248.91 & 230.63 \\
    B3LYP/6-31++G**   & 250.07 & 236.93 \\
    \bottomrule
  \end{tabular}
  \label{tab:tddft}
\end{table}

\noindent Quantitative evaluation of delocalisation time represents a significant challenge requiring the use of electron quantum dynamics methods such as the wave-packet propagation (WPP) approach\cite{Aguilar-Galindo2021, Gauyacq2004}. An important requirement for WPP methods is a high-level quantitative description of the excited state wavefunction, which plays a role of the initial condition for the Cauchy problem. This remains a largely unsolved, non-trivial task for  DFT methods\cite{Aguilar-Galindo2021}, which becomes even more complicated for larger chemical systems such as Ar@C$_{60}$ and Ar@C$_{60}$/Ag(111). Although MOM-DFT might, in principle, be a good source of the initial wavefunction required for the WPP calculations, implementing a theoretical framework to solve this long-standing problem goes far beyond the scope of the present paper.

\section{Relativistic considerations}
Although a relativistic treatment would further enhance the theoretical calculations, we found that it was difficult to include relativistic effects via TD-DFT in such a way that enables us to elucidate and distinguish the two transitions Ar~$2p_{3/2} \rightarrow 4s$ and  Ar~$2p_{1/2} \rightarrow 4s$. For example, spin--orbit coupling calculations in contemporary quantum chemistry packages such as ORCA tend to give coupling matrix elements between multiplets that arise from a non-relativistic TD-DFT calculation, and while this might be able to incorporate some relativistic effects into the computed excitation energies, we found that it was not straight-forward at all to obtain physical insights into the spin--orbit-coupled excitation energies that would give us the understanding or insights we sought.\\

\noindent Instead, we performed Dirac--Hartree--Fock (DHF) calculations\cite{Mohanty1991,Sun2021} including Breit and Gaunt interactions\cite{Breit1929,Malli2016} in a $j$-adapted Gaussian basis set as implemented in PySCF\cite{Sun2020}. As a single-determinantal method, DHF provides access to spin--orbit-coupled four-component molecular orbitals (bispinors) which allows us to identify the Ar $2p_{3/2}$ and $2p_{1/2}$ bispinors in the ground state of Ar@C60, easily. In fact, the DHF/6-31++G** calculation for the ground state yields doubly degenerate Ar $2p_{1/2}$ bispinors and quadruply degenerate Ar $2p_{3/2}$ bispinors that are separated by 2.13 eV, in good agreement with experimental results (Figure \ref{FigS2}).

\section{The Z+1 approximation: ground state K@C$_{60}$}
\begin{figure}[h!]
\centering
\includegraphics[scale=0.5]{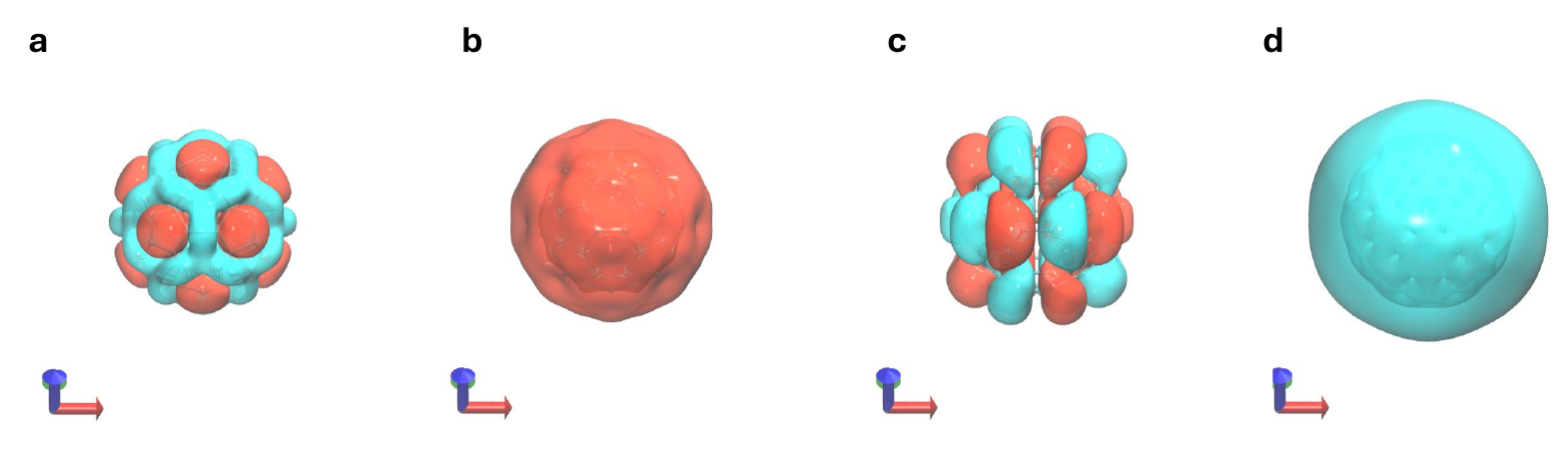}
\caption{\textbf{Kohn-Sham molecular orbitals for K@C$_{60}$, calculated at the PBE/6-31++G** level of theory.} All K contribution values were determined using the SCPA decomposition scheme discussed in Section 6 above. \textbf{(a)} Occupied, -15.08 eV, $A_g$, 53\% K; \textbf{(b)} occupied, -12.27 eV, $A_g$,60\% K; \textbf{(c)} HOMO, -4.27 eV, $T_{1u}$, 2\% K; \textbf{(d)} unoccupied, -1.45 eV, $A_g$, 98\% K 
}\label{FigS9}
\end{figure}

Key Kohn-Sham molecular orbitals for K@C$_{60}$ calculated at the PBE/~6-31++G** level of theory are shown in Fig. \ref{FigS9}. As noted in the main paper, while there is indeed some similarity between the Ar $2p \rightarrow 4s$ excited-state Ar@C$_{60}$ system and the ground state K@C$_{60}$ system by virtue of the $Z+1$ approximation, for the latter there exists both a first-order electron transfer from K to C$_{60}$ to generate the ion pair K$^+$(C$_{60}$)$^-$ and a second-order back-donation from (C$_{60}$)$^-$ to K$^+$ to reduce the charge separation in accordance with Pauling's principle of electroneutrality. On the other hand, in excited-state Ar@C$_{60}$ there exists only a partial charge transfer from Ar to C$_{60}$ via the distortion of the occupied Ar 4$s$ molecular orbital induced by the core hole, without any back-donation from the cage.

\section{Determining the Ar-Ag(111) separation: Supplementary density functional theory methods}
In addition to the LDA PAW-DFT results described in the main paper, we have carried out an extensive set of DFT calculations at various levels of theory in order to predict the separation of the encapsulated argon atom from the Ag(111) surface for the Ar@C$_{60}$ monolayer sample. The results are summarised in Table 3. Remarkably, we find best agreement with the experimentally measured (i.e. NIXSW derived) value of the Ar-Ag(111) separation from the most modest level of theory, i.e. LDA. Despite our expectation that good agreement would require careful implementation of dispersion force corrections, this surprisingly turned out to give substantially poorer agreement with experiment (as is clear from Table 3). A set of images of the various molecular adsorption configurations is available at the \href{http://doi.org/10.17639/nott.7457}{data repository} for this paper. 

\begin{table}[ht]
\centering
\arrayrulecolor{blue}
\begin{tabular}{|>{\columncolor{blue!10}}c|c|c|c|c|c|}
\hline
 & \textbf{bond-site} & \textbf{hex-hollow} & \textbf{hex-top} & \textbf{66 bond-hollow} & \textbf{66 bond-top} \\ \hline
\textbf{LDA} & 5.701 & 5.759 & 5.605 & 5.685 & \textcolor{teal}{5.549} \\ \hline
\textbf{PBE-D2} & 5.822 & 5.831 & 5.996 & 5.777 & 5.718 \\ \hline
\textbf{PBE-D3} & 5.847(!) & 5.839 & 5.770 & 5.820(!) & 5.740 \\ \hline
\textbf{SCAN} & - & - & - & - & 5.736 \\ \hline
\textbf{SCAN-D3} & - & - & - & - & 5.694 \\ \hline
\end{tabular}
\caption{The Ar-Ag separations (in \AA) calculated using different DFT functionals. The labels of the columns use the nomenclature molecular orientation-binding site. The experimentally determined value (see Fig. 3 of the main paper and associated discussion) is 5.54 $\pm$ 0.04 \AA.}
\label{tab:transposed_data}
\end{table}